\documentclass[pdflatex,sn-basic,dvipsnames]{sn-jnl}

\usepackage{natbib}
\usepackage{nicematrix}

\jyear{2021}%

\usepackage{amssymb}
\usepackage{amsthm}
\usepackage{amsmath}

\usepackage{multirow}
\usepackage{multicol}
\usepackage{subfig}
\usepackage{nicematrix}
\usepackage{makecell}
\newcolumntype{?}{!{\vrule width 2pt}}
\usepackage{pdfrender}
\newcommand*{\boldcheckmark}{%
  \textpdfrender{
    TextRenderingMode=FillStroke,
    LineWidth=.7pt, 
  }{\checkmark}%
}

\theoremstyle{thmstyleone}%
\theoremstyle{thmstyletwo}%

\theoremstyle{thmstylethree}%

\begin{document}

\title[Viability of Domain Constrained Collective Coalition Formation]{The Viability of Domain Constrained Coalition Formation for Robotic Collectives}


\author*[1]{\fnm{Grace} \sur{Diehl}}\email{diehlg@oregonstate.edu}

\author[1]{\fnm{Julie A.} \sur{Adams}}\email{julie.a.adams@oregonstate.edu}

\affil[1]{\orgdiv{Collaborative Robots and Intelligent Systems Institute}, \orgname{Oregon State University}, \orgaddress{
\city{Corvallis}, \postcode{97331}, \state{Oregon}, \country{United States}}}


\abstract{Applications, such as military and disaster response, can benefit from robotic collectives’ ability to perform multiple cooperative tasks (e.g., surveillance, damage assessments) efficiently across a large spatial area.  \textit{Coalition formation} algorithms can potentially facilitate collective robots' assignment to appropriate task teams; however, most coalition formation algorithms were designed for smaller multiple robot systems (i.e., 2-50 robots). Collectives' scale and domain-relevant constraints (i.e., distribution, near real-time, minimal communication) make coalition formation more challenging. This manuscript identifies the challenges inherent to designing coalition formation algorithms for very large collectives (e.g., 1000 robots). A survey of multiple robot coalition formation algorithms finds that most are unable to transfer directly to collectives, due to the identified system differences; however, auctions and hedonic games may be the most transferable. A simulation-based evaluation of three auction and hedonic game algorithms, applied to homogeneous and heterogeneous collectives, demonstrates that there are collective compositions for which no existing algorithm is viable; however, the experimental results and literature survey suggest paths forward.}




\maketitle

\section{Introduction}


Robotic collectives (i.e., $>50$ robots) are increasingly relevant for military (e.g., building or area surveillance) \citep{DefenseAdvancedResearchProjectsAgency2019OFFensiveTactics} and disaster response (e.g., damage assessments, search and rescue) \citep{Hildmann2019Review:Safety} applications. A fundamental problem is \textit{coalition formation for task allocation}, or partitioning robots into teams for task performance \citep{Gerkey2004ASystems,Korsah2013AAllocation}. Effective coalition formation can enable collectives to perform multiple cooperative tasks distributed over large environments efficiently; however, collectives' scale and the application domains' constraints make this problem challenging.

Military and disaster response applications require high-quality coalition formation solutions for hundreds to thousands of robots in near real-time (i.e., $<5$ minutes). Low-bandwidth, deployed networks (e.g., mobile ad hoc networks) will often be utilized, due to damaged permanent infrastructure or remote deployment locations \citep{Klinsompus2015CriticalPhases}. Minimal, distributed communication is required to reserve bandwidth for mission critical communications \citep{Jahir2019RoutingSurvey,Legendre201130Research,Shah2019TowardsNetworks}. Additionally, frequent communication with a central entity is infeasible, as collectives can be highly distributed \citep{Berman2009,Hamann2018SwarmApproach}.

Existing coalition formation incorporates exact and approximation  \citep[e.g.,][]{Aziz2021Multi-robotApproximation,Dutta2019One-to-manyAllocation, Service2011CoalitionAlgorithms, Zhang2013ConsideringAllocation}, %
auction-based \citep[e.g.,][]{Chen2011ResourceMethodology,Guerrero2017Multi-robotSolutions,Oh2017Market-BasedEnvironments}, 
biologically-inspired \citep[e.g.,][]{Agarwal2014Non-additiveFormation,Mouradian2017ADisasters,Yeh2016SolvingOptimization,Haque2013MultilevelMissions}, 
and, recently, hedonic game-based \citep[e.g.,][]{Czarnecki2019HedonicRobots,Jang2019AnRequirements} algorithms. However, software agent algorithms \citep[e.g.,][]{Liemhetcharat2014WeightedAgents,Michalak2010AGeneration,Rahwan2008AnGeneration,Rahwan2009AnGeneration, Shehory1995TaskAgents, Sless2014FormingNetworks} are not directly transferable to collectives due to differences between software agents and embodied robots \citep{Vig2007CoalitionRobots}. Additionally, robot coalition formation algorithms were evaluated predominantly for smaller multiple robot systems, not collectives, and few evaluations consider practical communication requirements, so it is unknown if the algorithms will scale. Thus, existing algorithms either are not, or have not been demonstrated to be, viable for collectives.

This manuscript assesses robot coalition formation algorithms' viability. A simulation-based evaluation with 
up to one thousand robots demonstrates that multiple robot coalition formation algorithms do not fully satisfy the target domains' solution quality, runtime, and communication requirements when applied to collectives. In fact, there are collective compositions for which no existing algorithm is suitable; however, potential avenues for addressing these collective compositions are suggested.

\subsection{Coalition Formation for Task Allocation}

Coalition formation for task allocation partitions agents into teams for task performance \citep{Service2011RandomizedGeneration}. The input is a set of $n$ agents, $A = \{a_1, \dots, a_n\}$, a set of $m$ tasks, $T = \{t_1, \dots, t_m\}$, and a \textit{characteristic function}, $f: (t_j, C_j) \rightarrow \mathbb{R}$, where $C_j \subseteq A$ is the set of agents assigned to task $t_j$ \citep{Service2011CoalitionAlgorithms}. Characteristic functions vary by application; 
however, $f(t_j, C_j)$ generally represents the inherent value of coalition $C_j$ completing task $t_j$  \citep{Service2011CoalitionAlgorithms,Zhang2013ConsideringAllocation}. 

A coalition formation for task allocation problem assumes that a coalition's value can be calculated considering only agents in the coalition (i.e., not considering agents assigned to other tasks) \citep{Rahwan2009AnGeneration}. 
The optimal solution is a \textit{coalition structure}, $CS = \{(t_1, C_1), \dots (t_m, C_m)\}$, that satisfies:

\begin{equation*}
    CS^* = \underset{CS}{\operatorname{argmax}} \sum_{j=1}^m f(t_j,C_j).
\end{equation*}

Deriving an optimal solution, or a provably reasonable approximation, is \textsc{NP}-complete \citep{Service2011CoalitionAlgorithms,Sandholm1999}. 
Specifically, no polynomial time algorithm can produce an $O(|C|^{1-\epsilon})$ or $O(m^{1-\epsilon})$ approximation, where $C$ is the set of non-zero valued coalitions and $\epsilon>0$, unless \textsc{P}$=$\textsc{NP} \citep{Service2011CoalitionAlgorithms, Sandholm1999}. Additionally, the establishment of any bound on solution quality requires considering at least $O(2^{n-1})$ coalition structures \citep{Sandholm1999}. This computational complexity hinders producing high-valued coalition structures for large numbers of agents.

\section{Coalition Formation for Multiple vs. Collective Robot Systems}

Coalition formation has received much attention in the software multi-agent and multiple robot 
communities, where a \textit{software multi-agent system} 
comprises two or more software agents, and a \textit{multiple robot system} is two to fifty robots. Software multi-agent coalition formation cannot be transferred directly to robotic 
systems, as robots have different constraints and practical considerations \citep{Vig2007CoalitionRobots}. As embodied agents, robots have 
kinematic and dynamic constraints 
and relatively static, nontransferable capabilities (i.e., sensors and actuators) \citep{Vig2007CoalitionRobots}. Robots are also 
likely to have power constraints \citep{Diehl2021BatterySwarms} and less available communication \citep{Diehl2021AnResponse}, which coalition formation must accommodate.

Multiple robot coalition formation considers embodied agents; however, transferring algorithms to robotic collectives can be complicated by differences in 
scale, capabilities, and communication. 
Key differences between collective and multiple robot systems must be considered when designing collective coalition formation algorithms, as summarized in Table \ref{Table:DomainComparison}.

\begin{table*}[t]
\begin{center}
\caption{Comparison of multiple robot and robotic collective systems.}
\label{Table:DomainComparison}
\begin{NiceTabular}{ |c||c|c| } 
 \hline
 & \textbf{Multiple Robot System} & \textbf{Robotic Collective} \\ 
 \hline
 \hline
 \textbf{Definition} &  2-50 robots & $> 50$ robots\\ 
 \hline
  \textbf{Agent} & \multirow{2}{*}{Sophisticated control}  & \multirow{2}{*}{Simple control} \\
  \textbf{Architecture} &&\\
 \hline
\textbf{System}  & Emergent behavior & Emergent behavior\\
\textbf{Capabilities} & possible & by design \\
 \hline
 \textbf{System} & Homogeneous or & Homogeneous or \\
  \textbf{Composition} &  heterogeneous & reduced heterogeneity\\
 \hline
  \multirow{2}{*}{\textbf{Communication}}  &  Fully connected network & Location-dependent\\ 
& Rich messages & Small, simple messages \\
 \hline
\end{NiceTabular}
\end{center}
\end{table*}

\subsection{Agent Architecture}

\label{sec:architecture}

Robotic collectives typically incorporate simple control models \citep{Brambilla2013} (e.g., repulsion, attraction, orientation  \citep{Hartman2006AutonomousBoids}), while multiple robot system architectures are often more sophisticated (e.g., belief-desire-intention). A reason for this difference is design choice. Robotic collectives were inspired by biological collectives, which achieve complex tasks through cooperation, rather than individual sophistication \citep{Brambilla2013}. Another reason is scale. Robotic collectives have more robots, so the allowable monetary and time costs per robot are lower; thus, even as robot technology and collective design improve, collective robots are likely to be less capable than their multiple robot system counterparts and have less computation available for coalition formation.

\subsection{System Capabilities}

\textit{System capabilities} are the tasks that a multiple robot system or collective can perform as a product of the individual robots' capabilities and interactions with other robots and the environment. Capabilities are relatively static when they correspond to hardware (e.g., sensors and actuators) \citep{Vig2007CoalitionRobots}, but \textit{emergent behaviors} can vary dynamically \citep{Beni2005}. 


Multiple robot systems can exhibit emergent behaviors, especially as more advanced artificial intelligence is incorporated (e.g., \cite{Costa2019OnlineTeams}); however, emergent behavior is intrinsic to collectives. Biologically-inspired collective capabilities (e.g., target selection \citep{Reina2015AMaking}) typically rely on emergent behaviors, and the use of reactive architectures (see Section \ref{sec:architecture}) renders interactions with the environment and other robots especially impactful \citep{Hartman2006AutonomousBoids}. Collective coalition formation must adapt task assignments quickly as emergent capabilities change.

\subsection{System Composition}
\label{sec:composition}

Collective and multiple robot systems may be \textit{homogeneous} (i.e., robots have identical capabilities) or \textit{heterogeneous} (i.e., some robots have different capabilities). Collective research has primarily considered homogeneous systems \citep[e.g.,][]{Reina2015AMaking,Prorok2017,VanDerBlom2018}, while multiple robot coalition formation has considered heterogeneity extensively \citep[e.g.,][]{Guerrero2017Multi-robotSolutions,Sen2013SA-ANT:Formation,Zhang2012IQ-ASyMTRe:Tasks}. Heterogeneous collectives enable more diverse applications
\citep{DefenseAdvancedResearchProjectsAgency2019OFFensiveTactics,Clark2021CCAST:Swarms,Prabhakar2020ErgodicAdaptation}; thus, collective coalition formation must also account for heterogeneity.

 Multiple robot systems can be highly heterogeneous, as each robot may be manufactured with unique capabilities \citep{Hamann2018SwarmApproach}, while collectives' larger scale means that robots are likely to be manufactured in batches, where each robot in a given batch has the same set of capabilities. Collective coalition formation algorithms may be able to leverage this reduced heterogeneity to decrease the required computation \citep{Service2011CoalitionAlgorithms}.

\subsection{Communication}

\label{sec:comms}

Coalition formation often assumes \textit{fully connected} communication networks in which each robot can communicate with every other robot directly (e.g., \cite{Jang2018AnonymousSystem,Tang2006ASyMTRe:Teams}). This assumption can be feasible for multiple robot systems, if the deployment domains have good network coverage \citep{Diehl2021BatterySwarms}; however, robotic collectives, like their biological counterparts, are likely to rely on distributed, local communication to facilitate scalability \citep{Berman2009,Hamann2018SwarmApproach}. Individuals in \textit{spatial swarms} (e.g., bird flocks \citep{Ballerini2008InteractionStudy}, fish schools \citep{Couzin2005EffectiveMove}) often communicate with only a subset of their neighbors \citep{Ballerini2008InteractionStudy,Haque2016AnalysisModels,Couzin2002CollectiveGroups,Strandburg-Peshkin2013VisualGroups}, while \textit{colonies} (e.g., bees \citep{Seeley2010HoneybeeDemocracy}, ants \citep{Gordon1999AntsOrganized}) generally have fully connected communication topologies within a central hub and no, limited, or spatial swarm-like, communication outside the hub \citep{Reina2015AMaking}. Spatial swarms and colonies can theoretically approximate fully connected topologies by using local communication to propagate messages, but at the cost of delays and solution quality degradation; thus, collective robot coalition formation needs to rely primarily on local communication.


Many collective and multiple robot system applications will use temporary, 
deployable communication networks (e.g., ad hoc networks), due to damaged permanent infrastructure or remote deployment locations \citep{Jahir2019RoutingSurvey,Legendre201130Research,Shah2019TowardsNetworks}. Temporary network nodes (e.g., ground and aerial robots, satellites, balloons, blimps, tablets, or smart phones \citep{Jahir2019RoutingSurvey,Pandey2017CommunicationChallenges})
cannot replace permanent infrastructure fully, given their comparatively limited power and bandwidth \citep{Muralidhar2018AnNetworks}. Thus, multiple robot systems may send only medium-sized, rich messages, while collective robots may send only small, simple messages, due to the collectives' scale (e.g., thousands of robots, large area coverage). 
Collective coalition formation will need to accommodate limited message sizes. 


\section{Background and Related Work}

Collective robots and coalition formation have, until recently \citep{Dutta2021DistributedAllocation, Czarnecki2021ScalableRobots}, been treated as distinct research areas, leaving collective coalition formation relatively unstudied. Collective and multiple robot coalition formation algorithms are discussed, as well as additional background on coalition formation.

\subsection{The Services Model}

Coalition formation requires determining if a coalition is capable of completing a task, given the coalition members' capabilities \citep{Vig2007CoalitionRobots}. There is a longstanding distinction between robot and software agent capability models, as hardware capabilities are not instantaneously transferable \citep{Vig2007CoalitionRobots}. Robots' software capabilities can also be more difficult to transfer, as software may be hardware-dependent (e.g., sensor-dependent controllers), and only limited communication may be available if a transfer is viable.

The robot \textit{services model} considers high-level robot behaviors (e.g., surveillance) called \textit{services} \citep{Service2011CoalitionAlgorithms,Vig2007CoalitionRobots}. Each robot has a service vector specifying the services that it can perform, and each task has a 
vector specifying the number of robots required to perform each service. A coalition 
can perform a task if there are sufficient robots to provide all required services, where each robot performs 
a single service at at time \citep{Service2011CoalitionAlgorithms}. This model is efficient when different sensors or actuators produce the same behavior, which supports scalability \citep{Service2011CoalitionAlgorithms}. Additionally, the services abstraction facilitates task design by users who are less familiar with robot hardware (e.g., tacticians \citep{DefenseAdvancedResearchProjectsAgency2019OFFensiveTactics}).

Three other common robot capabilities models exist. The \textit{robot types model}, which considers sets of robots with identical capabilities, 
can be converted to the services model \citep{Service2011CoalitionAlgorithms}. The \textit{resources model} considers tasks' resource (i.e., sensor and actuator) requirements and if resources are collocated or provided by different robots  \citep{Vig2007CoalitionRobots}. Certain problems (e.g., identifying the lowest cost coalition for a task) are more computationally difficult than with the services model \citep{Service2011CoalitionAlgorithms}, and more knowledge of robot hardware is needed for task design. The \textit{schema model} determines the information flow from robots' environmental sensors to code modules to actuators \citep{Tang2005ASyMTRe:Reconfiguration}. This model allows for very flexible task performance, but considers only motion-based (i.e., not processing or sensing) tasks. This manuscript focuses on applying the services model.

\subsection{Collective Coalition Formation}

Recent collective coalition formation leverages \textit{anonymous hedonic games}, where coalition size determines utility, and each robot tries to optimize its individual utility by joining and leaving coalitions until \textit{Nash stability} is reached (i.e., robots cannot benefit by individually changing coalitions) \citep{Dreze1980HedonicStability}. Centralized hedonic games \citep{Czarnecki2019HedonicRobots,Czarnecki2021ScalableRobots} are not well-suited to collectives, as they require frequent communication with a centralized computer (Section \ref{sec:comms}). Distributed algorithms are most relevant.

The distributed algorithms \citep{Jang2019AnRequirements,Jang2018AnonymousSystem,Dutta2021DistributedAllocation} derived from the GRoup Agent Partitioning and Placing Event (GRAPE) algorithm \citep{Jang2018AnonymousSystem}, in which each robot joins its preferred coalition, sends the resulting coalition structure to its network neighbors, and updates its coalition structure based on neighbors' messages. 
The GRAPE algorithm variations allow for homogeneous systems \citep{Jang2018AnonymousSystem}, variations in the performance of a single service \citep{Jang2019AnRequirements}, or heterogeneity under the resources model \citep{Dutta2021DistributedAllocation}; however, no services model variant exists.

The GRAPE variations are fast (i.e., milliseconds with 50 robots) and use local communication \citep{Jang2019AnRequirements,Jang2018AnonymousSystem,Dutta2021DistributedAllocation}; however, evaluation with large, heterogeneous collectives is an open problem. Additionally, individual robot utilities are not equivalent to coalition utilities, and differences may create suboptimal solutions \citep{Jang2019AnRequirements,Jang2018AnonymousSystem}. 

\subsection{Multiple Robot Coalition Formation}

Multiple robot coalition formation or \textit{ST-MR task allocation} (i.e., a robot performs a Single Task and a task is performed by Multiple Robots) is well-studied \citep{Gerkey2004ASystems,Korsah2013AAllocation}. The 
pros and cons of the primary algorithms are outlined.

\subsubsection{Exact and Approximation Algorithms}

Exact algorithms return optimal solutions, while approximation algorithms return solutions within a known factor of optimal. These solution quality guarantees can be advantageous; however, 
exact and approximation algorithms have not been favored historically for multiple robot coalition formation, due to coalition formation's high computational complexity \citep{Service2011CoalitionAlgorithms,Zhang2013ConsideringAllocation}.

These algorithm types are more feasible if the search space is constrained. Dynamic programming with a limited number of robot types, $j < O(n)$, produces optimal solutions with a $O(n^{2j}m)$ computational complexity 
\citep{Service2011CoalitionAlgorithms}, compared to the $O(3^n)$ 
otherwise required \citep{Rothkopf1998}. Limited robot types (see Section \ref{sec:composition}) can be a valid assumption; however, the dynamic programming algorithm is centralized, which is less compatible with collectives (see Section \ref{sec:comms}). 

Adaptations of Shehory and Kraus's distributed software multi-agent algorithm incorporate a maximum coalition size constraint, $k < O(n)$ \citep{Shehory1995TaskAgents, Shehory1998MethodsFormation}. The algorithms have two stages \citep{Vig2007CoalitionRobots, Vig2006Multi-robotFormation}. 
First, each robot calculates its possible coalitions' values. 
Next, robots broadcast their best coalitions. 
The best overall coalition is greedily added to the coalition structure until all tasks have been assigned or no coalitions can perform the remaining tasks. The base 
computational complexity is $O(n^km)$, and the approximation ratio depends on the heuristic used to identify the best coalition \citep{Shehory1995TaskAgents, Shehory1998MethodsFormation}.

There are three common heuristics. \textit{Maximum utility} selects the highest utility coalition and finds solutions within a factor $k+1$ of the optimal utility \citep{Service2011CoalitionAlgorithms}. \textit{Minimum cost} selects the lowest cost (i.e., 1/utility) coalition and finds solutions within a factor $O(k/\log k)$ of the optimal cost \citep{Shehory1998MethodsFormation}. \textit{Average utility} finds the coalition with the highest utility to size ratio, producing solutions within 
$2k$ of the optimal utility \citep{Zhang2013ConsideringAllocation}. This heuristic is useful when smaller coalitions are favored, and the utility does not depend on coalition size.

The \textit{resource-centric heuristic} aims to prevent robots that are needed for many tasks from being assigned early \citep{Zhang2013ConsideringAllocation}. The heuristic ranks coalitions based on utility and the degree to which their selection will limit future task allocation. The approximation factor is ($2k+2$), and the computational complexity is $O(min(n,m)m^2\binom{n}{k}^2)$ \citep{Zhang2013ConsideringAllocation}.

Any algorithm variation can incorporate the Fault Tolerance Coefficient to balance assigning redundant robots, in case some fail to perform tasks, with the cost of using extra robots \citep{Vig2006Multi-robotFormation}. Additionally, other algorithm variations exist. 
Service and Adams incorporated the services model and bipartite matching, reducing the computational complexity to $O(n^{3/2}m)$ \citep{Service2011CoalitionAlgorithms}. Dutta \textit{et al.}'s one-to-many bipartite matching-based algorithm has a $O(nm)$ computational complexity and a $(1/(1+max(k-\Delta)))$ cost approximation ratio, where $\Delta \in \{0, 1\}$; however, it only allows for homogeneous systems \citep{Dutta2019One-to-manyAllocation}.

The advantage of the greedy approximation algorithms is that they provide solution quality guarantees, are flexible in the objective functions that they can optimize, and facilitate heterogeneity. However, their reasonable computational complexities rely on a small maximum coalition size, $k$, which is unlikely given collectives' scale. The algorithms also rely on all-to-all broadcasts to communicate coalition values among the robots, which may be infeasible with deployed disaster response and military networks. Thus, existing variations 
cannot be easily applied to collective coalition formation.

\subsubsection{Auction Algorithms}

Auctions are a market-based resource allocation method, where buyers and sellers exchange information about the price at which they are willing to buy and sell goods \citep{Phelps2008AuctionsLearning}. A seller aims to obtain a high price, while a buyer's goal is to obtain goods for a low price. Auction algorithms are popular for multiple robot coalition formation, because they are an intuitive method of distributing resources. Existing algorithms vary based on the method of mapping tasks and robots to buyers and sellers, as well as the auction type.

Some algorithms have robots bid directly on tasks.  First-price, one round auction-based algorithms auction off tasks in the order that they are received \citep{Gerkey2002,Sujit2008UAVFormation}. An auctioneer broadcasts a task's required resources to the robots, robots reply with the cost of their participation, and the auctioneer chooses the lowest cost coalition that can perform the task. The advantage is fast runtimes relative to other auction mechanisms; however, a less valuable task received before a more valuable task is given precedence. Double auctions address this limitation by auctioning off all tasks at once \citep{Guerrero2017Multi-robotSolutions, Xie2018AFormation}. Robots  in Guerrero \textit{et al.}'s double auction algorithm send task auctioneers information about their relevant abilities, and task auctioneers send back bids based on the utility of the tasks they represent \citep{Guerrero2017Multi-robotSolutions}, while in Xie \textit{et al.}'s algorithm, the tasks start the auction process \citep{Xie2018AFormation}. These algorithms allow tasks to compete for resources, at the cost of additional communication.

Robots can also bid on tasks through intermediaries. Project Manager-Oriented Coalition Formation has robots elect project managers for each task, which select coalitions of their neighbors in the network topology \citep{Oh2017Market-BasedEnvironments}. This approach ensures that robots in a coalition can communicate, but 
severely restricts the solution space. The Automated Synthesis of Multi-robot Task solutions through software Reconfiguration (ASyMTRe) algorithm variants have robots form coalitions that bid on tasks \citep{Zhang2012IQ-ASyMTRe:Tasks,Tang2007AAllocation}. All ASyMTRe variations use the schema model. Finally, other algorithms have tasks bid on robot capabilities. The Robot Allocation through Coalitions Using Heterogeneous Non-Cooperative Agents (RACHNA) approach uses an ascending auction, with tasks bidding on services via service agents \citep{Vig2006Market-basedFormation}. A limitation is that robots can be unnecessarily reassigned. A simultaneous descending auction addresses this limitation, at the cost of an empirically higher runtime \citep{Service2014AAllocation}.

Overall, auction-based approaches are more applicable to collectives than exact and approximation algorithms, as they do not make assumptions about coalition size or require communication with a central computer. However,  the requirement that all robots communicate with the task auctioneers (or task agents with service agents) may be difficult for military and disaster response networks \citep{Vig2006Market-basedFormation,Service2014AAllocation}). The communication required is substantial and can increase at least linearly with 
the collective size, potentially exceeding deployed network's limited bandwidth. Additionally, algorithms were generally not evaluated with large collectives, so it is necessary to assess scalability.

\subsubsection{Biologically-Inspired Algorithms}

Early scalable 
robot coalition formation 
leveraged 
biologically-inspired optimization mechanisms. The simulated Annealing inspired ANT colony optimization (sA-ANT) algorithm, which leverages ant colony optimization and simulated annealing,
allocates only one task at a time \citep{Sen2013SA-ANT:Formation}. Double-layered ant colony optimization addresses this limitation \citep{Yeh2016SolvingOptimization}. Additionally, particle swarm optimization and the Pareto Archived Evolution Strategy-based algorithm addressed multi-objective coalition formation  \citep{Agarwal2014Non-additiveFormation,Mouradian2017ADisasters}.  

Most algorithms were evaluated in simulation at or near the scale of collectives \citep{Agarwal2014Non-additiveFormation, Mouradian2017ADisasters,Sen2013SA-ANT:Formation}; however, the algorithms are centralized and either allocate a single task at a time \citep{Mouradian2017ADisasters,Sen2013SA-ANT:Formation} or their experimental run times indicate requiring over 15 minutes for even small collectives \citep{Agarwal2014Non-additiveFormation,Yeh2016SolvingOptimization}. 

Haque \textit{et al.} proposed a decentralized algorithm based on alliance formation between male bottlenose dolphins \citep{Haque2013MultilevelMissions}; however, this algorithm restricts coalition sizes, which is not well suited to collectives.

\subsection{Summary}

The 
algorithm types most likely to be suitable for collectives are hedonic games and auctions. Hedonic games produce solutions quickly with multiple robot systems and can be distributed; however, distributed hedonic games have not been evaluated with large heterogeneous collectives (e.g., 1000 robots), and communication requirements have generally not been evaluated. Auctions are promising, given that some existing algorithms \citep[e.g.,][]{Zhang2012IQ-ASyMTRe:Tasks,Xie2018AFormation,Vig2006Market-basedFormation}) are decentralized and consider heterogeneous systems; however, auctions also have not been evaluated at the scale of collectives and may require excessive communication.

\section{Algorithms}

Hedonic game 
and auction 
coalition formation algorithms were evaluated for the viability of their use with collectives. Homogeneous GRAPE, a distributed hedonic game-based algorithm, is evaluated, as there is no services model variant \citep{Jang2018AnonymousSystem}. The distributed auction-based 
algorithms are RACHNA \citep{Vig2007CoalitionRobots, Vig2006Market-basedFormation} and simultaneous descending auction \citep{Service2014AAllocation}, which incorporate common auction protocols. 


\subsection{GRAPE}

\label{sec:grape}

GRAPE incorporates an 
anonymous hedonic game, meaning that each robot joins the most individually profitable coalition, as determined based on coalition size, not the coalition members' identities \citep{Jang2018AnonymousSystem}. During each iteration of GRAPE, robots select their highest valued coalition, broadcast their beliefs, about all robots' current task assignments, and update their belief states based on messages from neighboring robots in the network topology. A robot's message is given precedence if the robot's belief state has been updated more times, or the same number of times and more recently, than the receiving robot's. This 
precedence system serves as a \textit{distributed mutex} that allows only a single robot to alter the valid coalition assignments during each iteration. The algorithm completes when a Nash stable partition is derived.

GRAPE requires $O(m)$ computation and $O(n)$ communication per iteration on a single robot,  where $m$ is the number of tasks, and $n$ is the collective size \citep{Jang2018AnonymousSystem}. A Nash stable partition can be found in $O(n^2)$ iterations, given a fully connected network topology, resulting in $O(n^2m)$ computational and $O(n^3)$ communication complexities \citep{Jang2018AnonymousSystem}. Any connected network topology with a diameter $d_G$ has a $O(n^2md_G)$ computational complexity with a $O(n^3d_G)$ communication complexity \citep{Jang2018AnonymousSystem}. 

These complexities require that each robot's reward for any given task decreases as coalition size increases \citep{Jang2018AnonymousSystem}. This manuscript uses a \textit{peaked reward}, or a system reward that is highest when a task is assigned a coalition with exactly the desired number of robots, divided evenly among coalition members \citep{Jang2018AnonymousSystem}. An individual robot's reward for task $t_j$ with utility $u_j$ that requires $n_j$ robots and is assigned to coalition $C_j$ is:

\begin{equation}
    \label{eq:peaked_reward} 
    utility(t_j, C_j) = \frac{u_j}{n_j} \times e^{-\frac{\lvert C_j \rvert}{n_j}+1}.
\end{equation}

\noindent If multiple service types are available, this function cannot determine which robots will perform each service type; thus, GRAPE is considered only for homogeneous collectives with one service type.

\subsection{RACHNA}

RACHNA incorporates a \textit{combinatorial ascending auction}, meaning that buyers bid on bundles of goods, and bids increase as the auction progresses \citep{Vig2006Market-basedFormation}. The sellers are \textit{service agents} (i.e., one agent per service type), and the buyers are \textit{task agents}. 
Service agents sell a service type and track robots' current salaries (i.e., rewards coalition membership) \citep{Vig2006Market-basedFormation}. Task agents bid on service bundles, which they are awarded if the bid is at least the robots' salaries, plus $|S|\times \epsilon_{inc}$, where $|S|$ is the task's set of required services and $\epsilon_{inc}$ is a fixed wage increase. A tasks' maximum bid is its utility.

At most $n(u_{max}/\epsilon_{inc})$ bidding rounds occur, where $u_{max}$ is the highest task utility, with $O(m)$ bids per round. 
Task agents require $O(n)$ computation to determine their bids, and service agents require $O(n\log n)$ computation to accept/reject bids, resulting in $O(mn^2\log n (u_{max}/\epsilon_{inc}))$ total computation. A service agent communicates with at most all task agents and robots per round \citep{Vig2006Market-basedFormation}. Messages exchanged with task agents are size $O(n)$, and messages exchanged with the robots have size $O(1)$; thus, the communication complexity is $O(nm|S|)$ per round and $O(n^2m|S|(u_{max}/\epsilon_{inc}))$ in total.

Prior evaluation of RACHNA treated $\epsilon_{inc}$ as a fixed parameter controlling the degree of competition \citep{ Vig2007CoalitionRobots,Vig2006Market-basedFormation,Sen2013AFormation}. However, tasks for which $|S|\times\epsilon_{inc} > u_j$  cannot be assigned coalitions, even when sufficient robots 
exist. This limitation is incompatible with collectives' potentially large coalition sizes; thus, RACHNA must incorporate collective size into its $\epsilon_{inc}i$ threshold dynamically. \textit{RACHNA}$_{dt}$ denotes a new implementation of RACHNA where $\epsilon_{inc} = 1 / n$. 
RACHNA$_{dt}$'s computational complexity is $O(mn^3\log n (u_{max}))$, and the communication complexity is $O(n^3m|S|(u_{max}))$.

\subsection{Simultaneous Descending Auction}

Simultaneous descending auction, like RACHNA, incorporates a combinatorial auction with service and task agents. Each robots' salary is set initially to the maximum task utility, plus $\epsilon_{dec}$, where $\epsilon_{dec}$ is a fixed wage decrement. Robot's salaries are decremented by $\epsilon_{dec}$ at the beginning of each bidding round, and task agents that still require additional services bid on service bundles. If the task's utility is higher than the total salaries of all robots in the bundle, the task agent is able to afford a service bundle. The auction stops when all robots have been purchased or all salaries are zero.

Two 
implementations are considered. The first, simultaneous descending auction with a small coalition optimization (denoted SDA$_{SCO}$), has task agents determine their bids by enumerating all coalitions that meet their service requirements. The computational complexity is $O(mn^k)$ per iteration, where $k$ is the maximum coalition size \citep{Service2014AAllocation}. There are at most $u_{max}/\epsilon_{dec}$ iterations, resulting an an overall computational complexity of $O(mn^ku_{max}/\epsilon_{dec})$. This implementation is faster than the alternative with small coalitions, $k \leq 3$, and was evaluated in prior work \citep{Service2014AAllocation}.

The second implementation, denoted SDA$_M$, determines task agents' bids using weighted bipartite matching \citep{Service2014AAllocation}. The computational complexity is $O(mn^4u_{max}/\epsilon_{dec}$), regardless of coalition size \citep{Service2014AAllocation}. This complexity is identical to SA$_{SCO}$'s when $k=4$ and faster when $k>4$ \citep{Service2014AAllocation}. Both implementations have communication complexities of $O((m+n)|S|u_{max}/\epsilon_{dec})$, as service agents communicate with at most all robots and task agents during each iteration.

\section{Experimental Design}

A simulation-based experiment assessed each algorithm's \textit{viability}, where a viable algorithm produces near-optimal (i.e., $\geq95$\%) solutions in near real-time (e.g., $<5$ minutes) for a range of collective coalition formation problems. These criteria are representative of requirements for highly dynamic domains. Algorithms must additionally use as little communication as possible (i.e., $<500$ MB in total), so that a collectives' limited bandwidth may be used primarily for 
transmission of data relevant to task performance. 

The experiment considered 
\textit{achievable missions}, meaning that the collectives' robots offered sufficient services to perform all tasks simultaneously. Real-world collective deployments will ideally incorporate achievable missions but are unlikely to involve substantially more than the minimum required robots, due to the expense and logistical challenges of deploying large collectives. The experiment's collectives possessed exactly the minimum number of robots required 
(i.e., coalition formation solutions must utilize all robots).

The independent variables were the algorithm, collective size, number of tasks, and collective composition (see Table \ref{table:ind_variables}). The algorithms considered were GRAPE, RACHNA ($\epsilon_{inc}=1$), RACHNA$_{dt}$, SDA$_{SCO}$ ($\epsilon_{dec}=1$), and SDA$_M$ ($\epsilon_{dec}=1$). The \textit{collective size} varied from multiple robot systems (i.e., 25-50 robots) to collectives (i.e., 100-1000 robots). The numbers of tasks were 1\%, 10\%, and 50\% the number of robots, corresponding to average coalition sizes 100, 10, and 2, respectively. 1\% tasks (i.e., coalition size 100) was used only for collectives with $>$100 robots, as a mission is only  achievable if the number of robots per task is smaller than the collective size. Collective composition encompassed the number of \textit{service types} and \textit{services per robot}, where more capable robots had more services, and the number of 
service types and services per robot combinations determined the level of heterogeneity. A collective was homogeneous if the numbers of service types and services per robot were equal. Otherwise, the collective was heterogeneous.

\begin{table}[h]
    \centering
    \caption{Independent variables.}
    \label{table:ind_variables}
    \begin{NiceTabular}{|c|l|}
        \hline
        Algorithms & GRAPE, RACHNA, RACHNA$_{dt}$, SA$_{SCO}$ and SA$_M$\\
        \hline
        Collective Size & 25, 50, 100, 500, 1000 \\
        \hline
       Percent Tasks & 1\%, 10\%, 50\% (i.e., coalition sizes 100, 10, 2)\\
       \hline
        Service Types & 1, 5, 10\\
        \hline
        Services Per Robot & 1, 5\\
        \hline
    \end{NiceTabular}
  
\end{table}

GRAPE was considered only for homogeneous collectives with one service type, as GRAPE does not incorporate a services model. Additionally, SDA$_{SCO}$ was considered only for problems with multiple robot systems (i.e., 25 or 50 robots), due to the algorithm's high computational complexity.  
The other algorithms were considered for all independent variable value

The experiment used a centralized C++ simulator on a HP Z640 Workstation (Intel Xeon processor, 62 GB RAM) \citep{Sen2013AFormation}. The simulator performed each algorithm iteration for each robot sequentially and assumed a fully connected communication topology. The robots' embodiment was partially represented by the services model; however, the robots' positions were not considered, as incorporating positions in the utility function, when applicable, is highly dependent on the application environment and robot hardware. Twenty-five problem instances were randomly generated per independent variable combination, where a problem instance comprised sets of robots and tasks, each with associated services. Robots' and tasks' services were selected randomly, and each task was assigned a random integer utility in the range [1, 50]. Each trial, or problem, was allocated twelve hours to produce a solution. This time limit is too long for high tempo dynamic domains (e.g., disaster response) or even short term pre-mission planning (e.g., 2-4 hour breaks between mission deployments \citep{DefenseAdvancedResearchProjectsAgency2019OFFensiveTactics}).

The dependent variable \textit{success rate} represents the ratio of problems for which algorithms provided non-zero utility solutions within the time limit. A trial was \textit{unsuccessful} if the computer's memory limit was exceeded, the algorithm's runtime exceeded the 12-hour time limit, or the algorithm was otherwise unable to assign any robots to tasks. Only successful trials were considered when analyzing the other dependent variables, as unsuccessful trials can cause less suitable algorithms to appear to perform well under certain metrics. For example, an algorithm that quickly exceeds the memory limit, resulting in no viable solution, will have a very low runtime, but is less suitable than an algorithm that produces a solution given a longer runtime.

The other dependent variables are runtime, total communication, and percent utility. \textit{Runtime} is the time in minutes (min) and seconds (s) 
required for an algorithm to produce a solution. \textit{Total communication} is the sum of all message sizes in megabytes (MB). \textit{Percent utility} measures the solution quality of successful trials (i.e., solution utility/optimal utility). Overall, higher success rates and percent utilities with lower runtimes and total communication are preferred. It was \textit{hypothesized} that none of the algorithms will perform well for all metrics across the range of collective sizes and compositions.

\section{Results}

Results are presented for homogeneous and heterogeneous collectives. Box plots were used, as the data was not normally distributed. Non-paramet-ric statistical methods assessed significance across each independent variable. Mann-Whitney-Wilcoxon tests compared the numbers of service types and the services per robot, while Kruskal-Wallis analysis with Mann-Whitney-Wilcoxon post-hoc tests compared across the collective sizes and percent tasks. All analysis included only independent variable combinations for which an algorithm had successful trials across all independent variable values (e.g., percent task analysis considered only 500 and 1000 robot collectives). An algorithm was deemed viable if it had consistently low runtimes and communication, as well as high percent utilities, for all independent variable values.

\subsection{Homogeneous Collective Results}

A total of 600 trials with homogeneous collectives were run for each of RACHNA, RACHNA$_{dt}$, and SDA$_M$. Recall that GRAPE's analysis considered only the 300 trials with one service type, 
and $SDA_{SCO}$ was considered for only the 200 multiple robot trials. 

\subsubsection{GRAPE}

GRAPE produced optimal solutions for all trials with one service type, for a 50\% overall success rate. Recall that GRAPE was unable to perform the 300 trials with five service types, or half the trials. 

\begin{figure}[t!]
    \centering
    \includegraphics[width=\linewidth]{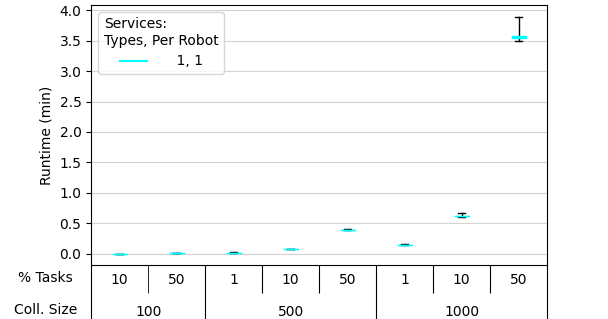}
    \caption{\textit{\textbf{GRAPE's runtimes (min) with homogeneous collectives.}}  GRAPE did not solve problems with 5 service types and 5 services per robot. All multiple robot trials (i.e., 25-50 robots) completed in $<$ 1 s. Note that the y-axis maximum is 4 min.
}
    \label{fig:grape_homogeneous_runtime}
\end{figure}

All successful trials had low runtimes, well within the $<5$ min target (Figure \ref{fig:grape_homogeneous_runtime}). Runtimes did increase substantially from multiple robot systems ($<1$ s) to collectives ($<3$ min 53 s), as well as with percent tasks. The increases were significant with collective size ($H$ ($n=50$) $=$ 238.94, $p < 0.01$) and percent tasks ($H$ ($n=50$) $=$ 74.50 $p < 0.01$), with significant post-hoc analyses ($p<0.01$). The longest runtimes were for 1000 robot collectives with 50\% tasks (i.e., the largest number of robots and tasks), consistent with GRAPE's $O(n^2m)$ computational complexity. Nevertheless, all of GRAPE's runtimes were sufficiently fast for high-tempo applications.


GRAPE's total communication increased only with collective size (Table \ref{tab:grape_comms}). This increase was significant ($H$ ($n=50$) $=$ 249.00, $p<0.01$), with significant post-hoc analyses ($p<0.01$).  Recall that GRAPE's per iteration communication complexity is $O(n)$. Percent tasks and problem instance did not impact the communication requirement, because all problems required $n$ iterations (i.e., no robots deviated from their initially selected coalitions).

\begin{table}[h!]
    \centering
     \caption{\textbf{GRAPE's communication (MB) for homogeneous collectives with one service type.} Communication was constant across problem instances with equal collectoive sizes and independent of other variables.}
    \begin{NiceTabular}{|c||c|}
    \hline
      Coll. Size   &  Communication (MB)\\
      \hline
     \hline
     25    & 1.87\\\hline
     50 & 7.5\\\hline
     100 & 30\\\hline
     500 & 750\\\hline
     1,000 & 3,000\\\hline
    \end{NiceTabular}
    \label{tab:grape_comms}
\end{table}

\subsubsection{RACHNA and \texorpdfstring{RACHNA$_{dt}$}{RACHNA\_dt}}

RACHNA$_{dt}$ solved all problems, while RACHNA produced solutions for only 82.3\% of trials. All RACHNA trials with 1\% tasks were unsuccessful, as well as 12\% with 25 robots, 10\% tasks, and either collective composition. All unsuccessful trials considered problems in which every tasks' required coalition size exceeded its utility, consistent with RACHNA's known limitation.

\begin{figure}[b]
\centering
\subfloat[\textbf{RACHNA}'s runtimes with homogeneous collectives.\label{subfig:rachna_runtime_homogeneous}]{%
      \includegraphics[width=0.95\linewidth]{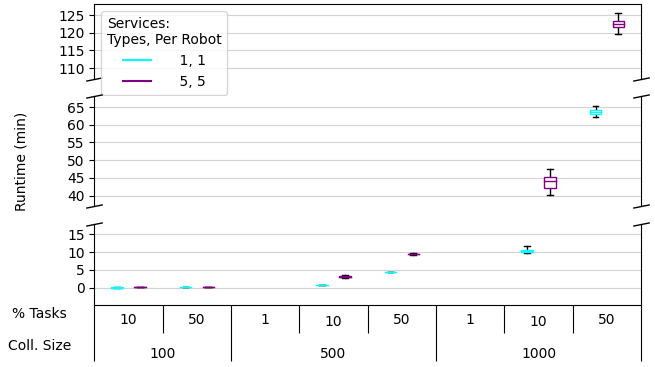}
    }
    \hfill
    \subfloat[\textbf{RACHNA$_{dt}$}'s runtimes with homogeneous collectives. \label{subfig:rachnadt_runtime_homogeneous}]{%
      \includegraphics[width=0.95\linewidth]{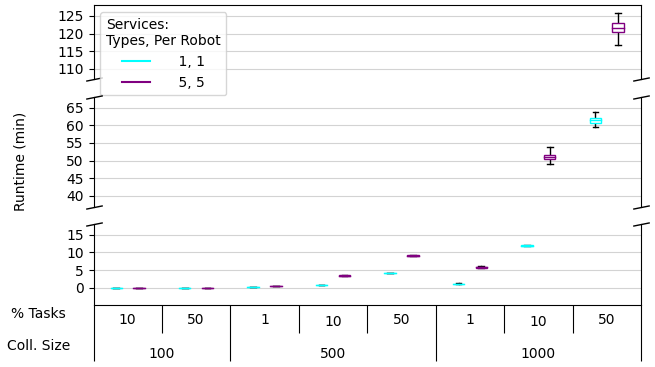}
    }
    
    \caption{\textit{\textbf{RACHNA's and RACHNA$_{dt}$'s runtimes (min) with homogeneous collectives.}} All multiple robot trials (i.e., 25-50 robots) completed in $<$ 1 s. RACHNA did not solve problems with 1\% tasks. The y-axis maximum is 125 min (i.e., 2 hours 5 min), and there are axis breaks.} 
    \label{fig:rachnas_homogeneous_runtime}
\end{figure}

RACHNA's and RACHNA$_{dt}$'s runtimes differed by $<1$ min (see Figure \ref{fig:rachnas_homogeneous_runtime}). Both algorithms had low runtimes with multiple robot systems ($<1$ s), which increased substantially for collectives ($<$ 125 min 48 s). Runtimes also increased with percent tasks, consistent with the 
$O(mn^3\log n)$ per iteration computational complexity. Additionally, runtimes increased with the number of service types, which corresponds to an increased number of service agents performing computation. The longest runtimes were for 1000 robots, 50\% tasks, and five service types. These runtimes were much too long for high-temp domains.

\begin{figure}[h!]
    \centering
    \subfloat[\textbf{RACHNA}'s communication with homogeneous collectives.\label{subfig:rachna_comms_homogeneous}]{%
      \includegraphics[width=0.92\linewidth]{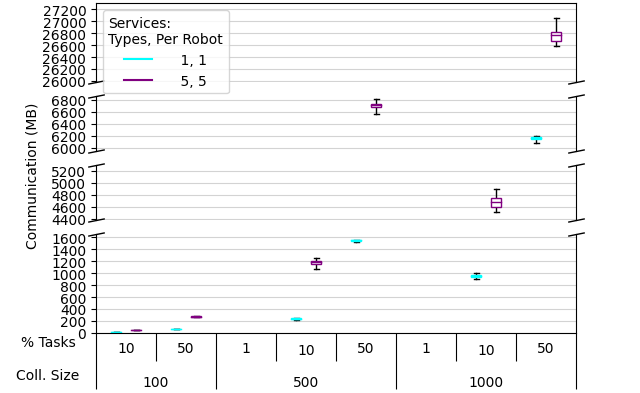}
    }
    \hfill
    \subfloat[\textbf{RACHNA$_{dt}$}'s communication with homogeneous collectives. \label{subfig:rachnadt_comms_homogeneous}]{%
      \includegraphics[width=0.92\linewidth]{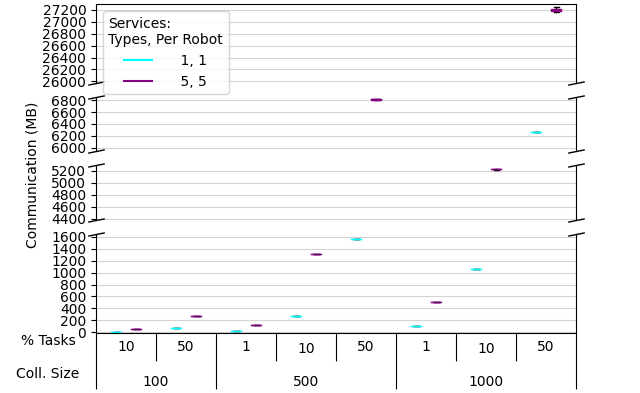}
    }
    \caption{\textit{\textbf{RACHNA's and RACHNA$_{dt}$'s communication (MB) with homogeneous collectives.}} All multiple robot trials (i.e., 25-50 robots) required $<70.81$ MB. RACHNA solved no problems with 1\% tasks. The y-axis maximum is 25,000 MB (25 GB).}
    \label{fig:rachna_rachnadt_homogeneous_comms}
\end{figure}

RACHNA's runtime increase with respect to collective size was significant ($H$ ($n_{25}=44, n=50$) $=$ 458.75, $p<0.01$), with significant post-hoc tests ($p<0.01$). Note that the lower number of samples for a collective size of 25 is due to unsuccessful trials.  RACHNA's runtime also increased significantly when the the percent tasks increased from 10\% to 50\% ($p<0.01$) and the service types increased from one to five $(p=0.01)$.

RACHNA$_{dt}$ produced similar results. RACHNA$_{dt}$'s runtimes increased significantly with collective size ($H$ ($n=50$) $=$ 467.32, $p<0.01$)) and percent tasks ($H$ ($n=50$) $=$ 138.08, $p<0.01$). Post-hoc tests found that all differences were significant ($p<0.01$). A significant difference also existed for increased numbers of service types ($p<0.01$).

RACHNA's and RACHNA$_{dt}$'s communication results also had similar trends (Figure \ref{fig:rachna_rachnadt_homogeneous_comms}), although RACHNA$_{dt}$ required up to an additional $600$ MB. Both algorithms' communication increased substantially from multiple robot systems ($\leq 70.81$ MB) to collectives ($\leq 27.25$ GB), as well as with percent tasks and service types. 1,000 robot collectives with 50\% tasks and five service types required the most communication, consistent with the algorithms $O(nm|S|)$ per iteration communication complexity. 

RACHNA's communication increase across collective sizes was significant ($H$ ($n_{25}=44, n=50$) $=$ 384.03, $p<0.01$), with significant post-hoc analyses ($p<0.01$). RACHNA's communication also increased significantly from 10\% to 50\% tasks ($P<0.01$) and one to five service types ($P<0.01$). Note that a Mann-Whitney-Wilcoxon test was used to assess significance with respect to percent tasks, as RACHNA had successful trials for only two values.

Similarly, RACHNA$_{dt}$'s communication increased significantly with collective size ($H$ ($n=50$) $=$ 390.97, $p<0.01$)) and percent tasks ($H$ ($n=50$) $=$ 234.59, $p<0.01$). Post-hoc tests identified significant differences between the collective sizes ($p<0.01$) and percent tasks ($p<0.01$). A significant increase from one to five service types was also identified ($p<0.01$).

RACHNA produced near-optimal solutions for most problems  (Table \ref{tab:e1_utility_rachna}), but high variability resulted in low worst case utilities. Unassigned tasks were generally those requiring more robots than supported by the task utility, given RACHNA's fixed threshold  limitation. Such tasks occur more often with lower percent tasks (i.e., larger coalitions) and more substantially impact mission performance with fewer robots (i.e., fewer tasks for a given value of \% tasks). Thus, multiple robot systems with 10\% tasks (i.e., the lowest percent utility with successful trials) had the most variable percent utilities.

Significant differences existed across the collective sizes ($H$ ($n_{25}=44, n=50$) $=$ 53.16, $p<0.01$), specifically, between 25 and 100-1000 robots ($p<0.01$), 50 and 100-1000 robots ($p=0.02$ for collective size 100, $p<0.01$, otherwise), and 100 and 1000 robots ($p<0.01$). No significant difference was found between 25 and 50 robots or 100 and 500 robots. 10\% and 50\% tasks also differed significantly ($p<0.01$), while one and five service types did not. 
RACHNA$_{dt}$ outperformed RACHNA, producing optimal solutions in all trials.

 \begin{table}[h!]
    \centering
    \caption{RACHNA's percent utility statistics with homogeneous collectives.}
    \begin{NiceTabular}{|c|c|c||c|}
    \hline
        Service & Percent & Collective &  Percent Utilities\\
         Types & Tasks & Size & Median (Minimum, Maximum)  \\
         \hline
         \hline
         \multirow{10}{*}{1}  & \multirow{5}{*}{10} & 25 &  100.0 (66.67, 100.0)\\
         && 50  & 99.17 (76.62, 100.0)\\
         && 100  & 96.77 (88.02, 100.0)\\
         && 500  & 96.39 (90.42, 99.040\\
         & & 1000 & 95.84 (92.94, 97.75) \\\cline{2-4}
        &\multirow{5}{*}{50} & 25  & 100.0 (98.91, 100.0)\\
         && 50  & 100.0 (99.6, 100.0)\\
         && 100  & 99.89 (99.56, 100.0)\\
         && 500  & 99.89 (99.8, 99.97)\\
         & & 1000  & 99.88 (99.76, 99.92)\\
         \hline
          \multirow{10}{*}{5} & \multirow{5}{*}{10} & 25  & 100.0 (72.73, 100.0)\\
         && 50  & 96.82 (80.91, 100.0)\\
         && 100  & 97.66 (90.17, 100.0)\\
         && 500  & 96.61 (92.88, 98.79)\\
         & & 1000  & 96.02 (94.02, 97.89)\\\cline{2-4}
           &  \multirow{5}{*}{50} & 25  & 100.0 (98.08, 100.0) \\
         && 50  & 99.93 (99.34, 100.0)\\
         && 100  & 99.84 (99.52, 100.0) \\
         && 500  & 99.91 (99.74, 100.0)\\
         & & 1000  & 99.89 (99.83, 99.95)\\
         
         \hline
    \end{NiceTabular}
    \label{tab:e1_utility_rachna}
\end{table}

\subsubsection{Simultaneous Descending Auction}

The results for the $SDA_M$ and $SDA_{SCO}$ simultaneous descending auction implementations are presented separately, due to vastly different performance. $SDA_M$ successfully produced optimal solutions in all trials.

\begin{figure}[t]
    \centering
    \includegraphics[width=\linewidth]{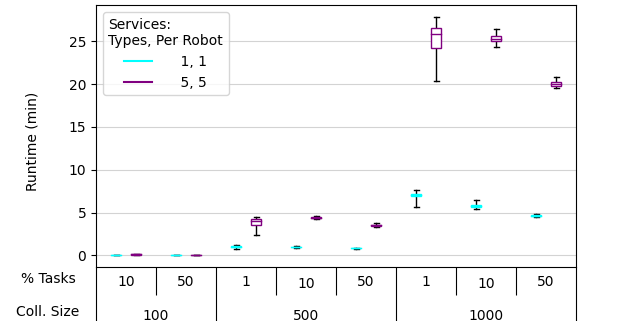}
    \caption{\textbf{$SA_M$'s runtimes (min) with homogeneous collectives.} All multiple robot trials (i.e., 25-50 robots) completed in $<1$ s. The y-axis maximum is 25 min.}
    \label{fig:samatch_homogeneous_runtime}
\end{figure}

$SDA_M$'s runtimes were within the $<5$ min target for most trials, but increased substantially for collectives with five service types (Figure \ref{fig:samatch_homogeneous_runtime}). Multiple robot trials had $<1$ s runtimes, which increased for collectives with one service type to $\leq$ 6 min 29 s. Collectives with five service types had runtimes $\leq$ 26 min 24 s. The increase with the number of service types can be attributed to the fact that task agents must consider each robot for more roles when determining their bids. Runtimes also decreased with the percent tasks, despite $SDA_M$'s $O(mn^4)$ per iteration computational complexity. This trend can be explained by the fact that higher percent tasks correspond to smaller coalitions. The smaller coalitions are cheaper for the task agents, thus enabling tasks to bid in fewer iterations. The longest runtimes occurred with 1000 robots, five service types and 50\% tasks.

The runtime increases across collective sizes ($H$ ($n=50$) $=$ 465.37, $p<0.01$) and tasks ($H$ ($n=50$) $=$ 10.61, $p<0.01$) were significant, as were the collective size post-hoc analyses ($p<0.01$); however, only differences between 10\% and 50\% tasks ($p<0.01$) and 1\% and 50\% tasks ($p=0.01$) were significant. No difference between 1\% and 10\% tasks was found. A significant increase from one to five service types was also identified ($p<0.01$).

\begin{figure}[t]
    \centering
     \subfloat[\textbf{Overview} of $SDA_M$'s communication.\label{subfig:samatch_comms_overview}]{%
      \includegraphics[width=0.9\linewidth]{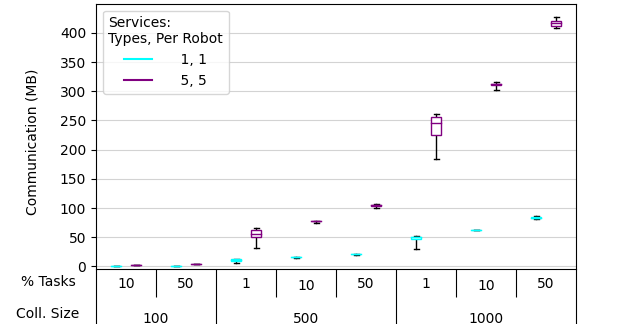}
      }
      \hfill
    
      \subfloat[Close-up with \textbf{1\% and 10\% tasks}.\label{subfig:samatch_comms_homogeneous_110}]{%
      \includegraphics[height=2.1in]{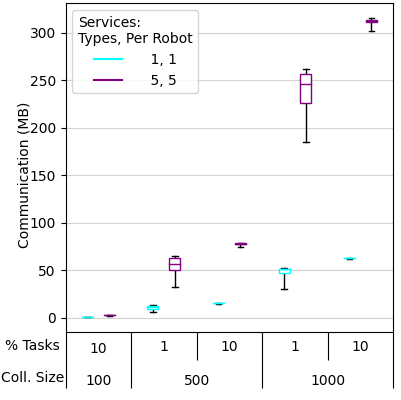}
    }
    ~ 
    \subfloat[Close-up with \textbf{50\% tasks}. \label{subfig:samatch_comms_homogeneous_50}]{%
      \includegraphics[height=2.1in]{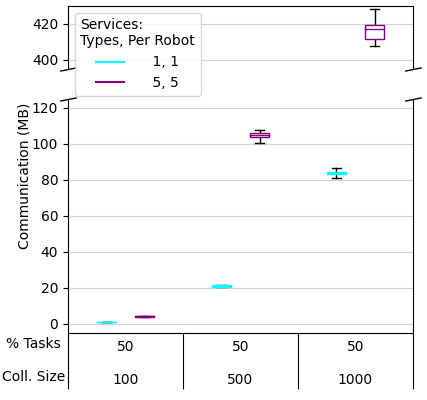}
      }
    \caption{\textbf{$SDA_M$'s communication (MB) with homogeneous collectives.} All multiple robot trials (i.e., 25-50 robots) required $<$ 1.2 MB. The y-axis varies between subfigures.}
    \label{fig:samatch_homogeneous_comms}
\end{figure}

$SDA_M$'s total communication was low with multiple robot systems ($<1.2$ MB), but increased substantially for robotic collectives (see Figure \ref{fig:samatch_homogeneous_comms}). Collectives with one service type used up to 86.29 MB total, while collectives with five service types used up to 428.03 MB. The total communication also increased with service types and  percent tasks, consistent with the $O((m+n)|S|)$ per iteration communication complexity. 1000 robot collectives with five service types and 50\% tasks required the most communication. 

Each of these trends was significant. Communication increased significantly across collective sizes ($H$ ($n=50$) $=$ 435.46, $p<0.01$) and tasks ($H$ ($n=50$) $=$ 35.17, $p<0.01$), with significant post-hoc analyses ($p<0.01$). A significant increase from one to five service types was also found ($p<0.01$).

$SDA_M$'s performance was generally better than $SDA_{SCO}$'s across all metrics. Recall that the $SDA_{SCO}$ evaluation considered only the 200 problem instances with multiple robot systems (33.3\% of all trials), due to $SDA_{SCO}$'s exponential computation requirement. $SDA_{SCO}$'s success rate with multiple robot systems was only 62.5\%, for an overall success rate of 20.8\%. 

The unsuccessful trials occurred when $SDA_{SCO}$ exceeded either the computer's memory limit or the 12-hour runtime limit. The memory limit was exceeded for the 25 trials with 50 robots, 10\% tasks, and one service type. The runtime limit was exceeded for the 50 trials with five service types and 10\% tasks, regardless of the number of robots. $SDA_{SCO}$ produced optimal solutions for the remaining 125 trials. 

 $SDA_{SCO}$'s runtimes increased with the number of robots, service types, and decreased percent tasks, consistent with $SDA_{SCO}$'s $O(mn^k)$ per iteration computational complexity (Table \ref{tab:sasco_homogeneous_runtime}). Note that decreased service types decreases $m$, but also increases $k$. Each of the differences was statistically significant ($p<0.01$). Runtimes were also more variable.

\begin{table}[h!]
    \centering
    \caption{\textbf{$SDA_{SCO}$'s runtimes with homogeneous multiple robot systems.} Collective trials (i.e., 100-1000 robots) were not attempted. No other trials were successful.}
    \begin{NiceTabular}{|c|c|c||c|}
    \hline
        Service & Percent & Collective &  Runtime (min:s:ms) \\
         Types & Tasks & Size & Median (Minimum, Maximum)  \\
         \hline
         \hline
         \multirow{3}{*}{1}  & \multirow{1}{*}{10} & 25 & 1:16:195  (0:08:040, 1:41:884)\\\cline{2-4}
        &\multirow{2}{*}{50} & 25 & 0:00:132 (0:00:72, 0:00:454)\\
         & & 50 & 0:03:141 (0:00:769, 1:45:969)\\
         \hline
          \multirow{2}{*}{5} & \multirow{2}{*}{50} & 25 & 0:01:918 (0:00:483, 8:41:294)\\
           && 50 & 1:4:711 (0:05:98, 400:34:373)\\
         \hline
    \end{NiceTabular}
    \label{tab:sasco_homogeneous_runtime}
\end{table}

The variation in runtimes was due to $SDA_{SCO}$'s sensitivity to differences in problem difficulty. $SDA_{SCO}$'s computational complexity is exponential in coalition size, and the longest runtimes for each independent variable combination were for the trials with the highest maximum coalition size. Variation in coalition size resulted in a worst case runtime (i.e., $>6$ hours), much too long for high tempo applications or short term pre-mission planning.

The communication increased with the number of robots, tasks, and service types (Table \ref{tab:sasco_homogeneous_comms}), consistent with $SDA_{SCO}$'s $O((m+n)|S|)$ communication complexity. Each of these increases was significant ($p<0.01$).

\begin{table}[h!]
    \centering
    \caption{\textbf{$SDA_{SCO}$'s communication with homogeneous multiple robot systems.} Collective trials (i.e., 100-1000 robots) were not attempted. No other trials were successful.}
    \begin{NiceTabular}{|c|c|c||c|}
    \hline
        Service & Percent & Collective &  Communication (MB) \\
         Types & Tasks & Size & Median (Minimum, Maximum)  \\
         \hline
         \hline
         \multirow{3}{*}{1}  & \multirow{1}{*}{10} & 25 & 25.25 (2.92, 36.96)\\\cline{2-4}
        &\multirow{2}{*}{50} & 25 & 48.25 (31.39, 56.74)\\
         & & 50 & 203.3 (165.96, 228.14)\\
         \hline
          \multirow{2}{*}{5} & \multirow{2}{*}{50} & 25 &239.83 (142.12, 280.6)\\
           && 50 & 1,001.37 (860.66, 1,109.07)\\
         \hline
    \end{NiceTabular}
    \label{tab:sasco_homogeneous_comms}
\end{table}

\subsection{Homogeneous Collective Discussion}

The homogeneous collective experiment assessed whether GRAPE, RAC-HNA, RACHNA$_{dt}$, $SDA_M$, and $SDA_{SCO}$ are suitable for very large homogeneous collectives in highly dynamic domains. The \textit{hypothesis} that no algorithm fully satisfied the 100\% success rate, $<5$ min runtime, $<500$ MB total communication, and $>95$\% utility criteria was supported.

$SDA_{SCO}$ satisfied the evaluation criteria for the fewest independent variable combinations (i.e., three out of twenty-four), as shown in Table \ref{tab:discussion_homogeneous_bad}. All criteria were satisfied for single-service systems with 10\% tasks and 25 robots, as well as 50\% tasks and 25-50 robots. Relaxing the target runtime to $<10$ min enables $SDA_{SCO}$ to satisfy all criteria for one additional independent variable combination (i.e., five services, 50\% tasks, 25 robots). This relaxation is relatively reasonable for near real-time coalition formation; however, there are no other independent variable combinations for which $SDA_{SCO}$ can satisfy all criteria, short of ignoring both the runtime and communication requirements entirely, which cannot be done in near real-time domains.

\begin{table}[h!]
    \centering
      \caption{\textbf{Homogeneous Results Summary: SDA$_{SCO}$ and RACHNA.} Each cell corresponds to a criterion for evaluating viability and an independent variable combination (i.e., an algorithm, services (S), percent tasks (T), and a collective size). A \textcolor{ForestGreen}{\boldcheckmark} means that the algorithm met the criterion for all trials with the independent variable combination, while a \textcolor{Apricot}{\boldcheckmark} means that all trials were reasonably close to meeting the criterion. An \textcolor{red}{\textbf{X}} means that the criterion was not met, or trials were not attempted due to known algorithm limitations. A \textbf{-} means that the independent variable combination is invalid. }
    \label{tab:discussion_homogeneous_bad}
    \begin{NiceTabular}{|c|c|c?c|c|c|c|c?c|c|c|c|c|}
    \hline
        & & & \multicolumn{5}{c?}{\textbf{SDA}$_{SCO}$} & \multicolumn{5}{c|}{\textbf{RACHNA}}\\\cline{4-13}
        & & & \multicolumn{5}{c?}{Collective Size}& \multicolumn{5}{c|}{Collective Size}\\\cline{4-13}
       & \multirow{-3}{*}{\makecell{S}} & \multirow{-3}{*}{\makecell{T}}  & 25 & 50 & 100 & 500 & 1000 & 25 & 50 & 100 & 500 & 1000\\
         \Xhline{2pt}
         \multirow{6}{*}{\makecell{Succ.\\Rate\\100\%}} &\multirow{3}{*}{1} & 1 &\textbf{-} & \textbf{-} & \textbf{-} & \textcolor{red}{\textbf{X}}& \textcolor{red}{\textbf{X}}&\textbf{-} & \textbf{-}&\textbf{-} & \textcolor{red}{\textbf{X}}& \textcolor{red}{\textbf{X}}\\\cline{3-13}
        & & 10 & \textcolor{ForestGreen}{\boldcheckmark} & \textcolor{red}{\textbf{X}}& \textcolor{red}{\textbf{X}} & \textcolor{red}{\textbf{X}}& \textcolor{red}{\textbf{X}}&\textcolor{red}{\textbf{X}} & \textcolor{ForestGreen}{\boldcheckmark}& \textcolor{ForestGreen}{\boldcheckmark} & \textcolor{ForestGreen}{\boldcheckmark}& \textcolor{ForestGreen}{\boldcheckmark}\\\cline{3-13}
        & & 50 & \textcolor{ForestGreen}{\boldcheckmark} & \textcolor{ForestGreen}{\boldcheckmark}& \textcolor{red}{\textbf{X}} & \textcolor{red}{\textbf{X}}& \textcolor{red}{\textbf{X}}&\textcolor{ForestGreen}{\boldcheckmark} & \textcolor{ForestGreen}{\boldcheckmark}& \textcolor{ForestGreen}{\boldcheckmark} & \textcolor{ForestGreen}{\boldcheckmark}& \textcolor{ForestGreen}{\boldcheckmark}\\\cline{2-13}
        
       & \multirow{3}{*}{5} & 1 & \textbf{-} & \textbf{-} & \textbf{-}& \textcolor{red}{\textbf{X}}& \textcolor{red}{\textbf{X}}& \textbf{-} & \textbf{-} & \textbf{-} & \textcolor{red}{\textbf{X}}& \textcolor{red}{\textbf{X}}\\\cline{3-13}
        & & 10 & \textcolor{red}{\textbf{X}} & \textcolor{red}{\textbf{X}}& \textcolor{red}{\textbf{X}} & \textcolor{red}{\textbf{X}}& \textcolor{red}{\textbf{X}}&  \textcolor{red}{\textbf{X}} &\textcolor{ForestGreen}{\boldcheckmark} & \textcolor{ForestGreen}{\boldcheckmark} & \textcolor{ForestGreen}{\boldcheckmark} & \textcolor{ForestGreen}{\boldcheckmark}\\\cline{3-13}
       & & 50 & \textcolor{ForestGreen}{\boldcheckmark} & \textcolor{ForestGreen}{\boldcheckmark}& \textcolor{red}{\textbf{X}} & \textcolor{red}{\textbf{X}}& \textcolor{red}{\textbf{X}} &\textcolor{ForestGreen}{\boldcheckmark} &\textcolor{ForestGreen}{\boldcheckmark} & \textcolor{ForestGreen}{\boldcheckmark} & \textcolor{ForestGreen}{\boldcheckmark} &\textcolor{ForestGreen}{\boldcheckmark}\\
        \Xhline{2pt}
         \multirow{6}{*}{\makecell{Run-\\time\\$<5$\\min}} &\multirow{3}{*}{1} & 1& \textbf{-} & \textbf{-} & \textbf{-} &\textcolor{red}{\textbf{X}} & \textcolor{red}{\textbf{X}} & 
         \textbf{-} & \textbf{-}&\textbf{-}& \textcolor{red}{\textbf{X}}& \textcolor{red}{\textbf{X}}\\\cline{3-13}
        & & 10 & \textcolor{ForestGreen}{\boldcheckmark} & \textcolor{red}{\textbf{X}}& \textcolor{red}{\textbf{X}} & \textcolor{red}{\textbf{X}}&\textcolor{red}{\textbf{X}}& \textcolor{ForestGreen}{\boldcheckmark} & \textcolor{ForestGreen}{\boldcheckmark}& \textcolor{ForestGreen}{\boldcheckmark} & \textcolor{ForestGreen}{\boldcheckmark} &  \textcolor{Apricot}{\boldcheckmark}\\\cline{3-13}
        & & 50 & \textcolor{ForestGreen}{\boldcheckmark} & \textcolor{ForestGreen}{\boldcheckmark}& \textcolor{red}{\textbf{X}} & \textcolor{red}{\textbf{X}}& \textcolor{red}{\textbf{X}}&\textcolor{ForestGreen}{\boldcheckmark} & \textcolor{ForestGreen}{\boldcheckmark}& \textcolor{ForestGreen}{\boldcheckmark} & \textcolor{ForestGreen}{\boldcheckmark}& \textcolor{red}{\textbf{X}}\\\cline{2-13}
       & \multirow{3}{*}{5} & 1 &\textbf{-} & \textbf{-} & \textbf{-} & \textcolor{red}{\textbf{X}}& \textcolor{red}{\textbf{X}}&\textbf{-} & \textbf{-} & \textbf{-} & \textcolor{red}{\textbf{X}} & \textcolor{red}{\textbf{X}}\\\cline{3-13}
        & & 10 & \textcolor{red}{\textbf{X}}&  \textcolor{red}{\textbf{X}}& \textcolor{red}{\textbf{X}}& \textcolor{red}{\textbf{X}}&  \textcolor{red}{\textbf{X}} &\textcolor{ForestGreen}{\boldcheckmark} & \textcolor{ForestGreen}{\boldcheckmark} &\textcolor{ForestGreen}{\boldcheckmark} & \textcolor{ForestGreen}{\boldcheckmark} & \textcolor{red}{\textbf{X}}\\\cline{3-13}
       & & 50 & \textcolor{Apricot}{\boldcheckmark}&  \textcolor{red}{\textbf{X}}& \textcolor{red}{\textbf{X}}& \textcolor{red}{\textbf{X}}&  \textcolor{red}{\textbf{X}} &\textcolor{ForestGreen}{\boldcheckmark} & \textcolor{ForestGreen}{\boldcheckmark} & \textcolor{ForestGreen}{\boldcheckmark} & \textcolor{Apricot}{\boldcheckmark}& \textcolor{red}{\textbf{X}}\\
        \Xhline{2pt}
          \multirow{6}{*}{\makecell{Comm.\\$<500$\\MB}} &\multirow{3}{*}{1} & 1 & \textbf{-} & \textbf{-} & \textbf{-} & \textcolor{red}{\textbf{X}} & \textcolor{red}{\textbf{X}} &\textbf{-} & \textbf{-} & \textbf{-} & \textcolor{red}{\textbf{X}} & \textcolor{red}{\textbf{X}}\\\cline{3-13}
        & & 10 & \textcolor{ForestGreen}{\boldcheckmark} &\textcolor{red}{\textbf{X}} & \textcolor{red}{\textbf{X}} & \textcolor{red}{\textbf{X}} & \textcolor{red}{\textbf{X}} &\textcolor{ForestGreen}{\boldcheckmark} & \textcolor{ForestGreen}{\boldcheckmark} & \textcolor{ForestGreen}{\boldcheckmark} & \textcolor{ForestGreen}{\boldcheckmark} & \textcolor{red}{\textbf{X}}\\\cline{3-13}
        & & 50 & \textcolor{ForestGreen}{\boldcheckmark} &\textcolor{ForestGreen}{\boldcheckmark} &\textcolor{red}{\textbf{X}}&\textcolor{red}{\textbf{X}}&\textcolor{red}{\textbf{X}} &\textcolor{ForestGreen}{\boldcheckmark} & \textcolor{ForestGreen}{\boldcheckmark} & \textcolor{ForestGreen}{\boldcheckmark} & \textcolor{red}{\textbf{X}} & \textcolor{red}{\textbf{X}}\\\cline{2-13}
       & \multirow{3}{*}{5} & 1 & \textbf{-} & \textbf{-} & \textbf{-} & \textcolor{red}{\textbf{X}} & \textcolor{red}{\textbf{X}} & \textbf{-} & \textbf{-} & \textbf{-} & \textcolor{red}{\textbf{X}} & \textcolor{red}{\textbf{X}}\\\cline{3-13}
        & & 10 &\textcolor{red}{\textbf{X}} &\textcolor{red}{\textbf{X}} &\textcolor{red}{\textbf{X}} &\textcolor{red}{\textbf{X}} &\textcolor{red}{\textbf{X}} &  \textcolor{ForestGreen}{\boldcheckmark} &\textcolor{ForestGreen}{\boldcheckmark} & \textcolor{ForestGreen}{\boldcheckmark} & \textcolor{red}{\textbf{X}} & \textcolor{red}{\textbf{X}}\\\cline{3-13}
       & & 50 & \textcolor{ForestGreen}{\boldcheckmark} &\textcolor{red}{\textbf{X}} & \textcolor{red}{\textbf{X}} &\textcolor{red}{\textbf{X}} &\textcolor{red}{\textbf{X}} &\textcolor{ForestGreen}{\boldcheckmark} & \textcolor{ForestGreen}{\boldcheckmark} & \textcolor{ForestGreen}{\boldcheckmark} & \textcolor{red}{\textbf{X}} & \textcolor{red}{\textbf{X}}\\
        \Xhline{2pt}
          \multirow{6}{*}{\makecell{Utility\\$>95$\%}} &\multirow{3}{*}{1} & 1& \textbf{-} & \textbf{-} & \textbf{-} & \textcolor{red}{\textbf{X}}& \textcolor{red}{\textbf{X}} & \textbf{-} & \textbf{-} & \textbf{-} & \textcolor{red}{\textbf{X}}& \textcolor{red}{\textbf{X}}\\\cline{3-13}
        & & 10 & \textcolor{ForestGreen}{\boldcheckmark} & \textcolor{red}{\textbf{X}}& \textcolor{red}{\textbf{X}} & \textcolor{red}{\textbf{X}}& \textcolor{red}{\textbf{X}} &\textcolor{red}{\textbf{X}} & \textcolor{red}{\textbf{X}} & \textcolor{red}{\textbf{X}} & \textcolor{Apricot}{\boldcheckmark}& \textcolor{Apricot}{\boldcheckmark} \\\cline{3-13}
        & & 50&  \textcolor{ForestGreen}{\boldcheckmark} & \textcolor{ForestGreen}{\boldcheckmark}& \textcolor{red}{\textbf{X}} & \textcolor{red}{\textbf{X}}& \textcolor{red}{\textbf{X}} &\textcolor{ForestGreen}{\boldcheckmark} & \textcolor{ForestGreen}{\boldcheckmark}& \textcolor{ForestGreen}{\boldcheckmark} & \textcolor{ForestGreen}{\boldcheckmark}& \textcolor{ForestGreen}{\boldcheckmark}\\\cline{2-13}
       & \multirow{3}{*}{5} & 1 & \textbf{-} & \textbf{-} & \textbf{-} & \textcolor{red}{\textbf{X}}& \textcolor{red}{\textbf{X}} &\textbf{-} & \textbf{-} & \textcolor{red}{\textbf{X}} & \textcolor{red}{\textbf{X}} & \textcolor{red}{\textbf{X}}\\\cline{3-13}
        & & 10 & \textcolor{red}{\textbf{X}} & \textcolor{red}{\textbf{X}}& \textcolor{red}{\textbf{X}} & \textcolor{red}{\textbf{X}}& \textcolor{red}{\textbf{X}} &\textcolor{red}{\textbf{X}} & \textcolor{red}{\textbf{X}} & \textcolor{Apricot}{\boldcheckmark}& \textcolor{Apricot}{\boldcheckmark}& \textcolor{Apricot}{\boldcheckmark} \\\cline{3-13}
       & & 50 & \textcolor{ForestGreen}{\boldcheckmark} & \textcolor{ForestGreen}{\boldcheckmark}& \textcolor{red}{\textbf{X}} & \textcolor{red}{\textbf{X}}& \textcolor{red}{\textbf{X}} &\textcolor{ForestGreen}{\boldcheckmark} & \textcolor{ForestGreen}{\boldcheckmark}& \textcolor{ForestGreen}{\boldcheckmark} & \textcolor{ForestGreen}{\boldcheckmark}& \textcolor{ForestGreen}{\boldcheckmark}\\
        \hline
        
    \end{NiceTabular}
  
\end{table}

$SDA_{SCO}$'s evaluation considered only multiple robot systems (i.e., $\leq50$ robots) due to a high computational complexity; however, it is clear from $SDA_{SCO}$'s poor performance with 25 and 50 robots that the algorithm will not scale to collectives. $SDA_{SCO}$'s major limitation is the excessive memory and computation required to enumerate the possible coalitions for each task. The evaluation criteria were satisfied only when there were few possible coalitions (i.e., 25 robots, 50\% tasks, 1 service type), while performance degraded substantially with higher numbers of possible coalitions (i.e., increased service types, increased numbers of robots, decreased percent tasks). Homogeneous collectives are expected to possess large numbers of robots, as well as be assigned tasks that require large coalitions (e.g., assessing damage to a city block), which will result in even worse performance than occurred with multiple robot systems. Thus, $SDA_{SCO}$ is not a viable approach for near real-time coalition formation with homogeneous collectives.

The next least viable algorithm was RACHNA (Table \ref{tab:discussion_homogeneous_bad}). RACHNA fully satisfied the evaluation criteria for only six independent variable combinations (i.e., one or five services with 50\% tasks and 25-100 robots). Relaxing the runtime requirement to $<10$ min and the utility requirement to $>90$\% enables RACHNA to satisfy all requirements for three additional independent variable combinations (i.e., one service with 10\% tasks and 500 robots, five services with 10\% tasks and 100 robots, five services with 50\% tasks and 500 robots). However, RACHNA cannot satisfy all evaluation criteria for the remaining independent variable combinations due to low percent utilities, as well as high worst-case runtimes and communication.

One of RACHNA's limitations was its high runtimes and communication requirements with 500-1000 robots, which well exceeded the evaluation criteria. The other major limitation was RACHNA's inability to assign coalitions to tasks with lower utilities than required coalition sizes. This limitation prevented RACHNA from solving any coalition formation problems with 1\% tasks (i.e., 100 robot coalitions), as well as satisfying the utility criterion with 10\% tasks (i.e., 10 robot coalitions). The limitation was masked somewhat for 10\% tasks with increased collective size, due to the increased number of tasks. However, the percent utilities and success rates were lower than RACHNA$_{dt}$'s, which, paired with comparable communication and runtimes, means that RACHNA$_{dt}$ is generally preferable to RACHNA.

\begin{table}[h!]
    \centering
      \caption{\textbf{Homogeneous Results Summary:  RACHNA$_{dt}$ and GRAPE.} Each cell corresponds to a criterion for evaluating viability and an independent variable combination (i.e., an algorithm, services (S), percent tasks (T), and a collective size). A \textcolor{ForestGreen}{\boldcheckmark} means that the algorithm met the criterion for all trials with the independent variable combination, while a \textcolor{Apricot}{\boldcheckmark} means that all trials were reasonably close to meeting the criterion. An \textcolor{red}{\textbf{X}} means that the criterion was not met, or trials were not attempted due to known algorithm limitations. A \textbf{-} means that the independent variable combination is invalid.  }
    \label{tab:discussion_homogeneous_good}
    \begin{NiceTabular}{|c|c|c?c|c|c|c|c?c|c|c|c|c|}
    \hline
        & & & \multicolumn{5}{c?}{\textbf{RACHNA$_{dt}$}} & \multicolumn{5}{c|}{\textbf{GRAPE}}\\\cline{4-13}
        & & & \multicolumn{5}{c?}{Collective Size}& \multicolumn{5}{c|}{Collective Size}\\\cline{4-13}
       & \multirow{-3}{*}{\makecell{S}} & \multirow{-3}{*}{\makecell{T}}  & 25 & 50 & 100 & 500 & 1000 & 25 & 50 & 100 & 500 & 1000\\
        \Xhline{2pt}
         \multirow{6}{*}{\makecell{Succ.\\Rate\\100\%}} &\multirow{3}{*}{1} & 1 & \textbf{-} & \textbf{-} & \textbf{-} & \textcolor{ForestGreen}{\boldcheckmark}& \textcolor{ForestGreen}{\boldcheckmark}&\textbf{-} & \textbf{-}& \textbf{-} & \textcolor{ForestGreen}{\boldcheckmark}& \textcolor{ForestGreen}{\boldcheckmark}\\\cline{3-13}
        & & 10 & \textcolor{ForestGreen}{\boldcheckmark} & \textcolor{ForestGreen}{\boldcheckmark}& \textcolor{ForestGreen}{\boldcheckmark} & \textcolor{ForestGreen}{\boldcheckmark}& \textcolor{ForestGreen}{\boldcheckmark}&\textcolor{ForestGreen}{\boldcheckmark} & \textcolor{ForestGreen}{\boldcheckmark}& \textcolor{ForestGreen}{\boldcheckmark} & \textcolor{ForestGreen}{\boldcheckmark}& \textcolor{ForestGreen}{\boldcheckmark}\\\cline{3-13}
        & & 50 & \textcolor{ForestGreen}{\boldcheckmark} & \textcolor{ForestGreen}{\boldcheckmark}& \textcolor{ForestGreen}{\boldcheckmark} & \textcolor{ForestGreen}{\boldcheckmark}& \textcolor{ForestGreen}{\boldcheckmark}&\textcolor{ForestGreen}{\boldcheckmark} & \textcolor{ForestGreen}{\boldcheckmark}& \textcolor{ForestGreen}{\boldcheckmark} & \textcolor{ForestGreen}{\boldcheckmark}& \textcolor{ForestGreen}{\boldcheckmark}\\\cline{2-13}
       & \multirow{3}{*}{5} & 1 & \textbf{-} & \textbf{-} & \textbf{-}& \textcolor{ForestGreen}{\boldcheckmark}& \textcolor{ForestGreen}{\boldcheckmark}& \textbf{-} & \textbf{-} & \textbf{-} & \textcolor{red}{\textbf{X}} & \textcolor{red}{\textbf{X}}\\\cline{3-13}
        & & 10 & \textcolor{ForestGreen}{\boldcheckmark} & \textcolor{ForestGreen}{\boldcheckmark}& \textcolor{ForestGreen}{\boldcheckmark} & \textcolor{ForestGreen}{\boldcheckmark}& \textcolor{ForestGreen}{\boldcheckmark}&  \textcolor{red}{\textbf{X}} & \textcolor{red}{\textbf{X}} & \textcolor{red}{\textbf{X}} & \textcolor{red}{\textbf{X}} & \textcolor{red}{\textbf{X}}\\\cline{3-13}
       & & 50 & \textcolor{ForestGreen}{\boldcheckmark} & \textcolor{ForestGreen}{\boldcheckmark}& \textcolor{ForestGreen}{\boldcheckmark} & \textcolor{ForestGreen}{\boldcheckmark}& \textcolor{ForestGreen}{\boldcheckmark} & \textcolor{red}{\textbf{X}} & \textcolor{red}{\textbf{X}} & \textcolor{red}{\textbf{X}} & \textcolor{red}{\textbf{X}} & \textcolor{red}{\textbf{X}}\\
        \Xhline{2pt}
         \multirow{6}{*}{\makecell{Run-\\time\\$<5$\\min}} &\multirow{3}{*}{1} & 1& \textbf{-} & \textbf{-} & \textbf{-} &\textcolor{ForestGreen}{\boldcheckmark} & \textcolor{ForestGreen}{\boldcheckmark}  & 
         \textbf{-} & \textbf{-}& \textbf{-}& \textcolor{ForestGreen}{\boldcheckmark}& \textcolor{ForestGreen}{\boldcheckmark}\\\cline{3-13}
        & & 10 & \textcolor{ForestGreen}{\boldcheckmark} & \textcolor{ForestGreen}{\boldcheckmark}& \textcolor{ForestGreen}{\boldcheckmark} & \textcolor{ForestGreen}{\boldcheckmark}& \textcolor{Apricot}{\boldcheckmark} &
        \textcolor{ForestGreen}{\boldcheckmark} & \textcolor{ForestGreen}{\boldcheckmark}& \textcolor{ForestGreen}{\boldcheckmark} & \textcolor{ForestGreen}{\boldcheckmark} &  \textcolor{ForestGreen}{\boldcheckmark}\\\cline{3-13}
        & & 50 & \textcolor{ForestGreen}{\boldcheckmark} & \textcolor{ForestGreen}{\boldcheckmark}& \textcolor{ForestGreen}{\boldcheckmark} & \textcolor{ForestGreen}{\boldcheckmark}& \textcolor{red}{\textbf{X}}&
        \textcolor{ForestGreen}{\boldcheckmark} & \textcolor{ForestGreen}{\boldcheckmark}& \textcolor{ForestGreen}{\boldcheckmark} & \textcolor{ForestGreen}{\boldcheckmark}& \textcolor{ForestGreen}{\boldcheckmark}\\\cline{2-13}
       & \multirow{3}{*}{5} & 1 &\textbf{-} & \textbf{-} & \textbf{-} & \textcolor{ForestGreen}{\boldcheckmark}& \textcolor{Apricot}{\boldcheckmark}&\textbf{-} & \textbf{-} & \textbf{-} & \textcolor{red}{\textbf{X}} & \textcolor{red}{\textbf{X}}\\\cline{3-13}
        & & 10 & \textcolor{ForestGreen}{\boldcheckmark}&  \textcolor{ForestGreen}{\boldcheckmark}& \textcolor{ForestGreen}{\boldcheckmark}& \textcolor{ForestGreen}{\boldcheckmark}&  \textcolor{red}{\textbf{X}} & \textcolor{red}{\textbf{X}} & \textcolor{red}{\textbf{X}} & \textcolor{red}{\textbf{X}} & \textcolor{red}{\textbf{X}} & \textcolor{red}{\textbf{X}}\\\cline{3-13}
       & & 50 & \textcolor{ForestGreen}{\boldcheckmark}&  \textcolor{ForestGreen}{\boldcheckmark}& \textcolor{ForestGreen}{\boldcheckmark}& \textcolor{Apricot}{\boldcheckmark}&  \textcolor{red}{\textbf{X}} & \textcolor{red}{\textbf{X}} & \textcolor{red}{\textbf{X}} & \textcolor{red}{\textbf{X}} & \textcolor{red}{\textbf{X}} & \textcolor{red}{\textbf{X}}\\
       \Xhline{2pt}
          \multirow{6}{*}{\makecell{Comm.\\$<500$\\MB}} &\multirow{3}{*}{1} & 1 & \textbf{-} & \textbf{-} & \textbf{-} & \textcolor{ForestGreen}{\boldcheckmark} & \textcolor{ForestGreen}{\boldcheckmark} &\textbf{-} & \textbf{-} & \textbf{-} & \textcolor{red}{\textbf{X}} & \textcolor{red}{\textbf{X}}\\\cline{3-13}
        & & 10 & \textcolor{ForestGreen}{\boldcheckmark} &\textcolor{ForestGreen}{\boldcheckmark} &\textcolor{ForestGreen}{\boldcheckmark} &\textcolor{ForestGreen}{\boldcheckmark} &\textcolor{red}{\textbf{X}} &\textcolor{ForestGreen}{\boldcheckmark} & \textcolor{ForestGreen}{\boldcheckmark} & \textcolor{ForestGreen}{\boldcheckmark} & \textcolor{red}{\textbf{X}} & \textcolor{red}{\textbf{X}}\\\cline{3-13}
        & & 50 & \textcolor{ForestGreen}{\boldcheckmark} &\textcolor{ForestGreen}{\boldcheckmark} &\textcolor{ForestGreen}{\boldcheckmark} &\textcolor{red}{\textbf{X}} &\textcolor{red}{\textbf{X}} &\textcolor{ForestGreen}{\boldcheckmark} & \textcolor{ForestGreen}{\boldcheckmark} & \textcolor{ForestGreen}{\boldcheckmark} & \textcolor{red}{\textbf{X}} & \textcolor{red}{\textbf{X}}\\\cline{2-13}
       & \multirow{3}{*}{5} & 1 & \textbf{-} & \textbf{-} & \textbf{-} & \textcolor{ForestGreen}{\boldcheckmark}  & \textcolor{Apricot}{\boldcheckmark} & \textbf{-} & \textbf{-} & \textbf{-} & \textcolor{red}{\textbf{X}} & \textcolor{red}{\textbf{X}}\\\cline{3-13}
        & & 10 &\textcolor{ForestGreen}{\boldcheckmark} &\textcolor{ForestGreen}{\boldcheckmark} &\textcolor{ForestGreen}{\boldcheckmark} &\textcolor{red}{\textbf{X}} &\textcolor{red}{\textbf{X}} &  \textcolor{red}{\textbf{X}} & \textcolor{red}{\textbf{X}} & \textcolor{red}{\textbf{X}} & \textcolor{red}{\textbf{X}} & \textcolor{red}{\textbf{X}}\\\cline{3-13}
       & & 50 & \textcolor{ForestGreen}{\boldcheckmark} &\textcolor{ForestGreen}{\boldcheckmark} &\textcolor{ForestGreen}{\boldcheckmark} &\textcolor{red}{\textbf{X}} &\textcolor{red}{\textbf{X}} &\textcolor{red}{\textbf{X}} & \textcolor{red}{\textbf{X}} & \textcolor{red}{\textbf{X}} & \textcolor{red}{\textbf{X}} & \textcolor{red}{\textbf{X}}\\
        \Xhline{2pt}
          \multirow{6}{*}{\makecell{Utility\\$>95$\%}} &\multirow{3}{*}{1} & 1& \textbf{-} & \textbf{-} & \textbf{-} & \textcolor{ForestGreen}{\boldcheckmark}& \textcolor{ForestGreen}{\boldcheckmark} & \textbf{-} & \textbf{-} & \textbf{-} & \textcolor{ForestGreen}{\boldcheckmark}& \textcolor{ForestGreen}{\boldcheckmark}\\\cline{3-13}
        & & 10 & \textcolor{ForestGreen}{\boldcheckmark} & \textcolor{ForestGreen}{\boldcheckmark}& \textcolor{ForestGreen}{\boldcheckmark} & \textcolor{ForestGreen}{\boldcheckmark}& \textcolor{ForestGreen}{\boldcheckmark} &\textcolor{ForestGreen}{\boldcheckmark} & \textcolor{ForestGreen}{\boldcheckmark}& \textcolor{ForestGreen}{\boldcheckmark} & \textcolor{ForestGreen}{\boldcheckmark}& \textcolor{ForestGreen}{\boldcheckmark} \\\cline{3-13}
        & & 50&  \textcolor{ForestGreen}{\boldcheckmark} & \textcolor{ForestGreen}{\boldcheckmark}& \textcolor{ForestGreen}{\boldcheckmark} & \textcolor{ForestGreen}{\boldcheckmark}& \textcolor{ForestGreen}{\boldcheckmark} &\textcolor{ForestGreen}{\boldcheckmark} & \textcolor{ForestGreen}{\boldcheckmark}& \textcolor{ForestGreen}{\boldcheckmark} & \textcolor{ForestGreen}{\boldcheckmark}& \textcolor{ForestGreen}{\boldcheckmark}\\\cline{2-13}
       & \multirow{3}{*}{5} & 1 & \textbf{-} & \textbf{-} & \textbf{-} & \textcolor{ForestGreen}{\boldcheckmark}& \textcolor{ForestGreen}{\boldcheckmark} &\textbf{-} & \textbf{-} & \textbf{-} & \textcolor{red}{\textbf{X}} & \textcolor{red}{\textbf{X}}\\\cline{3-13}
        & & 10 &  \textcolor{ForestGreen}{\boldcheckmark} & \textcolor{ForestGreen}{\boldcheckmark}& \textcolor{ForestGreen}{\boldcheckmark} & \textcolor{ForestGreen}{\boldcheckmark}& \textcolor{ForestGreen}{\boldcheckmark} &\textcolor{red}{\textbf{X}} & \textcolor{red}{\textbf{X}} & \textcolor{red}{\textbf{X}} & \textcolor{red}{\textbf{X}} & \textcolor{red}{\textbf{X}}\\\cline{3-13}
       & & 50 & \textcolor{ForestGreen}{\boldcheckmark} & \textcolor{ForestGreen}{\boldcheckmark}& \textcolor{ForestGreen}{\boldcheckmark} & \textcolor{ForestGreen}{\boldcheckmark}& \textcolor{ForestGreen}{\boldcheckmark} &\textcolor{red}{\textbf{X}} & \textcolor{red}{\textbf{X}} & \textcolor{red}{\textbf{X}} & \textcolor{red}{\textbf{X}} & \textcolor{red}{\textbf{X}}\\
        \hline
        
    \end{NiceTabular}
  
\end{table}

RACHNA$_{dt}$ satisfied all evaluation criteria for sixteen independent variable combinations, including all combinations with multiple robot systems, as well as 100 robots (see Table \ref{tab:discussion_homogeneous_good}). All criteria were also satisfied for single-service collectives with 500 robots and 1-10\% tasks, as well as 1000 robots and 1\% tasks, and five service collectives with 500 robots and 1\% tasks. Relaxing the communication requirement to $<600$ MB enables RACHNA$_{dt}$ to satisfy the criteria for one additional independent variable combination (i.e., five services, 1\% tasks, 100 robots). RACHNA$_{dt}$'s $>30$ min runtimes and $>1$ GB communication with all other independent variable combinations far exceeded the acceptable limits for near real-time domains.

RACHNA$_{dt}$'s primary strength was that the success rate and utility criteria were always met, given sufficient time and communication. This performance makes RACHNA$_{dt}$ well-suited for applications in which pre-mission planning is possible (i.e., the tasks are known at least a few hours prior to deployment). However, RACHNA$_{dt}$'s runtime and communication with 500-1000 robots are not suitable for near real-time domains, as the runtime and communication requirements are only met for some percent tasks (i.e., coalition sizes) and number of services. The number of services will generally be known prior to deployment, but the percent tasks will not, as tasks arise dynamically; thus, it generally cannot be guaranteed that RACHNA$_{dt}$ will be able to produce solutions within the runtime and communication requirements during any deployment with very large homogeneous collectives.

GRAPE fully satisfied the evaluation criteria for fewer independent variable combinations than RACHNA$_{dt}$ (i.e., the six combinations with one service and 25-100 robots), as shown in Table \ref{tab:discussion_homogeneous_good}. GRAPE's primary limitation was its inability to solve any coalition formation problems with multiple service types, due to its lack of a services model. However, GRAPE with only one service type satisfied the success rate, runtime, and percent utility requirements for all independent variable combinations. The number of service types will generally be known prior to deployment, so GRAPE, unlike RACHNA$_{dt}$, is suitable for deployment in near real-time domains, given single service collectives and sufficient communication infrastructure. Deployed networks will require reducing GRAPE's communication complexity, while multiple service collectives will require integration with a services model.

$SDA_M$ best satisfied the evaluation criteria, meeting all criteria for nineteen independent variable combinations (see Table \ref{tab:discussion_homogeneous_best}). The success rate, communication, and utility criteria were satisfied for all independent variable combinations. The runtime requirement was also met with 25-500 robots, and relaxing the requirement to $<10$ min enables $SDA_M$ to satisfy the criteria with 1000 robots and a single service type. However, $SDA_M$'s runtimes with five service types and 1000 robots were too long for near real-time domains.

Overall, $SDA_M$ satisfied the domain criteria reasonably well for single service collectives, and GRAPE produced solutions even more quickly, given sufficient communication infrastructure. However, no algorithm fully satisfied the performance criteria with five service types and 1000 robots. 

\begin{table}[h!]
    \centering
      \caption{\textbf{Homogeneous Results Summary: SDA$_M$.} Each cell corresponds to a criterion for evaluating viability and an independent variable combination (i.e., services (S), percent tasks (T), and a collective size). A \textcolor{ForestGreen}{\boldcheckmark} means that SDA$_M$ met the criterion for all trials with the independent variable combination, while a \textcolor{Apricot}{\boldcheckmark} means that all trials were reasonably close to meeting the criterion. An \textcolor{red}{\textbf{X}} means that the criterion was not met. A \textbf{-} means that the independent variable combination is invalid.}
    \label{tab:discussion_homogeneous_best}
    \begin{NiceTabular}{|c|c|c?c|c|c|c|c|}
        \hline
         & & & \multicolumn{5}{c|}{\textbf{SDA}$_{M}$}\\\cline{4-8}
          & & & \multicolumn{5}{c|}{Collective Size}\\\cline{4-8}
          & \multirow{-3}{*}{\makecell{S}} & \multirow{-3}{*}{\makecell{T}}  & 25 & 50 & 100 & 500 & 1000\\
          \Xhline{2pt}
          \multirow{6}{*}{\makecell{Succ.\\Rate\\100\%}} &\multirow{3}{*}{1} & 1 & \textbf{-} & \textbf{-} &\textbf{-} & \textcolor{ForestGreen}{\boldcheckmark}& \textcolor{ForestGreen}{\boldcheckmark}\\\cline{3-8}
          & & 10 & \textcolor{ForestGreen}{\boldcheckmark} & \textcolor{ForestGreen}{\boldcheckmark}& \textcolor{ForestGreen}{\boldcheckmark} & \textcolor{ForestGreen}{\boldcheckmark}& \textcolor{ForestGreen}{\boldcheckmark}\\\cline{3-8}
          & & 50 & \textcolor{ForestGreen}{\boldcheckmark} & \textcolor{ForestGreen}{\boldcheckmark}& \textcolor{ForestGreen}{\boldcheckmark} & \textcolor{ForestGreen}{\boldcheckmark}& \textcolor{ForestGreen}{\boldcheckmark}\\\cline{2-8}
           & \multirow{3}{*}{5} & 1 & \textbf{-} & \textbf{-} & \textbf{-}& \textcolor{ForestGreen}{\boldcheckmark}& \textcolor{ForestGreen}{\boldcheckmark}\\\cline{3-8}
           & & 10 & \textcolor{ForestGreen}{\boldcheckmark} & \textcolor{ForestGreen}{\boldcheckmark}& \textcolor{ForestGreen}{\boldcheckmark} & \textcolor{ForestGreen}{\boldcheckmark}& \textcolor{ForestGreen}{\boldcheckmark}\\\cline{3-8}
       & & 50 & \textcolor{ForestGreen}{\boldcheckmark} & \textcolor{ForestGreen}{\boldcheckmark}& \textcolor{ForestGreen}{\boldcheckmark} & \textcolor{ForestGreen}{\boldcheckmark}& \textcolor{ForestGreen}{\boldcheckmark}\\
        \Xhline{1.5pt}
        \multirow{6}{*}{\makecell{Run-\\time\\$<5$\\min}} &\multirow{3}{*}{1} & 1& \textbf{-} & \textbf{-} &\textbf{ -} &\textcolor{ForestGreen}{\boldcheckmark} & \textcolor{Apricot}{\boldcheckmark}\\\cline{3-8}
        & & 10 & \textcolor{ForestGreen}{\boldcheckmark} & \textcolor{ForestGreen}{\boldcheckmark}& \textcolor{ForestGreen}{\boldcheckmark} & \textcolor{ForestGreen}{\boldcheckmark}& \textcolor{Apricot}{\boldcheckmark}\\\cline{3-8}
        & & 50 & \textcolor{ForestGreen}{\boldcheckmark} & \textcolor{ForestGreen}{\boldcheckmark}& \textcolor{ForestGreen}{\boldcheckmark} & \textcolor{ForestGreen}{\boldcheckmark}& \textcolor{ForestGreen}{\boldcheckmark}\\\cline{2-8}
       & \multirow{3}{*}{5} & 1 &\textbf{-} &\textbf{ -} & \textbf{-} & \textcolor{ForestGreen}{\boldcheckmark}& \textcolor{red}{\textbf{X}}\\\cline{3-8}
        & & 10 & \textcolor{ForestGreen}{\boldcheckmark}&  \textcolor{ForestGreen}{\boldcheckmark}& \textcolor{ForestGreen}{\boldcheckmark}& \textcolor{ForestGreen}{\boldcheckmark}&  \textcolor{red}{\textbf{X}}\\\cline{3-8}
       & & 50 & \textcolor{ForestGreen}{\boldcheckmark}&  \textcolor{ForestGreen}{\boldcheckmark}& \textcolor{ForestGreen}{\boldcheckmark}& \textcolor{ForestGreen}{\boldcheckmark}&  \textcolor{red}{\textbf{X}}\\
        \Xhline{2pt}
         \multirow{6}{*}{\makecell{Comm.\\$<500$\\MB}} &\multirow{3}{*}{1} & 1 & \textbf{-} & \textbf{-} & \textbf{-} & \textcolor{ForestGreen}{\boldcheckmark} & \textcolor{ForestGreen}{\boldcheckmark}\\\cline{3-8}
        & & 10 & \textcolor{ForestGreen}{\boldcheckmark} &\textcolor{ForestGreen}{\boldcheckmark} &\textcolor{ForestGreen}{\boldcheckmark} &\textcolor{ForestGreen}{\boldcheckmark} &\textcolor{ForestGreen}{\boldcheckmark}\\\cline{3-8}
        & & 50 & \textcolor{ForestGreen}{\boldcheckmark} &\textcolor{ForestGreen}{\boldcheckmark} &\textcolor{ForestGreen}{\boldcheckmark} &\textcolor{ForestGreen}{\boldcheckmark} &\textcolor{ForestGreen}{\boldcheckmark}\\\cline{2-8}
       & \multirow{3}{*}{5} & 1 &\textbf{ -} & \textbf{-} & \textbf{-} & \textcolor{ForestGreen}{\boldcheckmark}  & \textcolor{ForestGreen}{\boldcheckmark}\\\cline{3-8}
        & & 10 &\textcolor{ForestGreen}{\boldcheckmark} &\textcolor{ForestGreen}{\boldcheckmark} &\textcolor{ForestGreen}{\boldcheckmark} &\textcolor{ForestGreen}{\boldcheckmark} &\textcolor{ForestGreen}{\boldcheckmark}\\\cline{3-8}
       & & 50 & \textcolor{ForestGreen}{\boldcheckmark} &\textcolor{ForestGreen}{\boldcheckmark} &\textcolor{ForestGreen}{\boldcheckmark} &\textcolor{ForestGreen}{\boldcheckmark} &\textcolor{ForestGreen}{\boldcheckmark}\\
        \Xhline{2pt}
          \multirow{6}{*}{\makecell{Utility\\$>95$\%}} &\multirow{3}{*}{1} & 1& \textbf{-} & \textbf{-} & \textbf{-} & \textcolor{ForestGreen}{\boldcheckmark}& \textcolor{ForestGreen}{\boldcheckmark}\\\cline{3-8}
        & & 10 & \textcolor{ForestGreen}{\boldcheckmark} & \textcolor{ForestGreen}{\boldcheckmark}& \textcolor{ForestGreen}{\boldcheckmark} & \textcolor{ForestGreen}{\boldcheckmark}& \textcolor{ForestGreen}{\boldcheckmark} \\\cline{3-8}
        & & 50&  \textcolor{ForestGreen}{\boldcheckmark} & \textcolor{ForestGreen}{\boldcheckmark}& \textcolor{ForestGreen}{\boldcheckmark} & \textcolor{ForestGreen}{\boldcheckmark}& \textcolor{ForestGreen}{\boldcheckmark}\\\cline{2-8}
       & \multirow{3}{*}{5} & 1 & \textbf{-} & \textbf{-} & \textbf{-} & \textcolor{ForestGreen}{\boldcheckmark}& \textcolor{ForestGreen}{\boldcheckmark}\\\cline{3-8}
        & & 10 &  \textcolor{ForestGreen}{\boldcheckmark} & \textcolor{ForestGreen}{\boldcheckmark}& \textcolor{ForestGreen}{\boldcheckmark} & \textcolor{ForestGreen}{\boldcheckmark}& \textcolor{ForestGreen}{\boldcheckmark}\\\cline{3-8}
       & & 50 & \textcolor{ForestGreen}{\boldcheckmark} & \textcolor{ForestGreen}{\boldcheckmark}& \textcolor{ForestGreen}{\boldcheckmark} & \textcolor{ForestGreen}{\boldcheckmark}& \textcolor{ForestGreen}{\boldcheckmark}\\
        \hline
    \end{NiceTabular}
\end{table}

\subsection{Heterogeneous Collective Results}

A total of 900 heterogeneous collective trials were run for each of RACH-NA, RACHNA$_{dt}$, and $SDA_{M}$. $SDA_{SCO}$ was run for only the 300 trials with multiple robot systems (i.e., 25-50 robots), due to its exponential computational complexity. Recall that GRAPE was not considered for heterogeneous collectives, as it does not incorporate the services model. 

\begin{figure}[b]
\centering
\subfloat[\textbf{RACHNA}'s runtimes with heterogeneous collectives.\label{subfig:rachna_runtime_heterogeneous}]{%
      \includegraphics[width=0.95\linewidth]{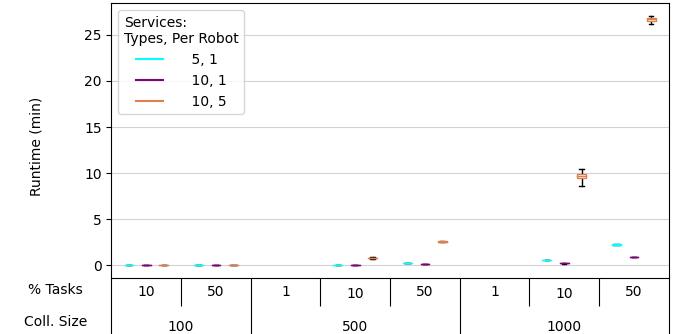}
    }
    \hfill
    \subfloat[\textbf{RACHNA$_{dt}$}'s runtimes with heterogeneous collectives. \label{subfig:rachnadt_runtime_heterogeneous}]{%
      \includegraphics[width=0.95\linewidth]{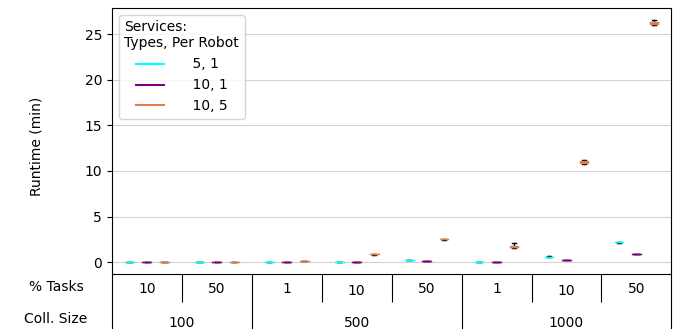}
    }
    
    \caption{\textit{\textbf{RACHNA's and RACHNA$_{dt}$'s runtimes (min) with heterogeneous collectives.}} All multiple robot trials (i.e., 25-50 robots) completed in $<$ 1 s. RACHNA did not solve problems with 1\% tasks. Note that the y-axis maximum is 25 min.}
    \label{fig:rachna_rachnadt_heterogeneous_runtime}
\end{figure}

\subsubsection{RACHNA and \texorpdfstring{RACHNA$_{dt}$}{RACHNA\_dt}}

RACHNA produced solutions for only 82.9\% of trials. The 150 trials with 1\% tasks were unsuccessful, due to RACHNA's known limitation. The other four unsuccessful trials occurred with 25 robots and 10\% tasks. Specifically, there was one unsuccessful trial with five services and one service per robot, two trials with ten services and one service per robot, and one trial with ten services and five services per robot. RACHNA$_{dt}$ had a 100\% success rate.

RACHNA and RACHNA$_{dt}$'s runtimes differed by $\leq 6$ min, as shown in Figure \ref{fig:rachna_rachnadt_heterogeneous_runtime}. Both algorithms' runtimes  were low for multiple robot systems ($<1$ s) and increased substantially for collectives with ten service types and five services per robot ($\leq27$ min 5 s). The runtimes were lower with five service types and one service per robot ($\leq$ 2 min 16 s), as well as ten service types and one service per robot ($\leq$ 53 s). Runtimes also increased with percent tasks, consistent with the algorithms' $O(mn^2\log n)$ per iteration computational complexity. Additionally, the runtimes increased with services per robot and decreased service types, as the number of services per robot divided by the service types determines the number of robots that each service agent must consider. The longest runtimes occurred with 1000 robots, ten service types, five services per robot, and 50\% tasks.

RACHNA's runtime increase with respect to collective size was significant ($H$ ($n_{25}=71$, $n=75$) $=$ 627.11, $p<0.01$), with significant post-hoc analyses ($p<0.01$). RACHNA's runtime also increased significantly with percent tasks ($p<0.01$), decreased service types ($p=0.02$), and services per robot ($p<0.01$). Note that service type analysis compared five service types and one service per robot to ten service types and one service per robot. Similarly, services per robot analysis compared ten service types and one service per robot to ten service types and five services per robot.

RACHNA$_{dt}$'s runtimes increased significantly across collective sizes ($H$ ($n=75$) $=$ 637.45, $p<0.01$) and tasks ($H$ ($n=150$) $=$ 150.44, $p<0.01$), with significant post-hoc analyses ($p<0.01$). Significant differences between services ($p<0.01$) and services per robot ($p=0.02$) were also identified.

\begin{figure}[h!]
\centering
\subfloat[\textbf{RACHNA}'s communication with heterogeneous collectives.\label{subfig:rachna_comms_heterogeneous}]{%
      \includegraphics[width=0.82\linewidth]{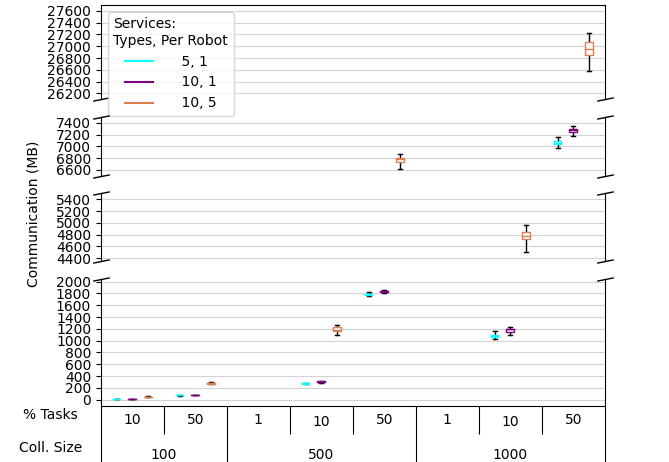}
    }
    \hfill
    \subfloat[\textbf{RACHNA$_{dt}$}'s communication with heterogeneous collectives. \label{subfig:rachnadt_commms_heterogeneous}]{%
      \includegraphics[width=0.82\linewidth]{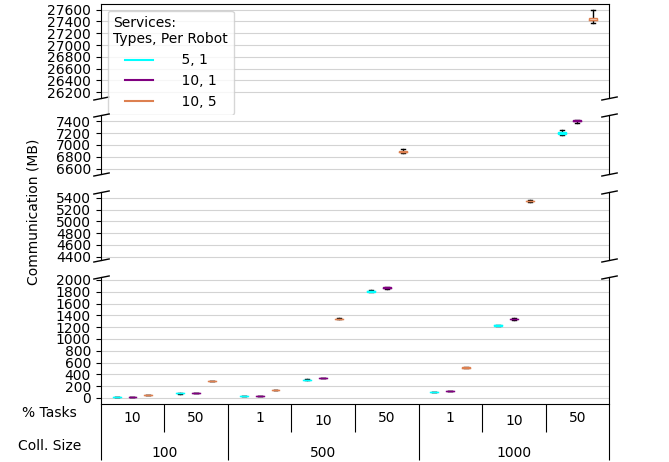}
    }
    
    \caption{\textit{\textbf{RACHNA's and RACHNA$_{dt}$'s communication (MB) with heterogeneous collectives.}} All multiple robot trials (i.e., 25-50 robots) required $<$ 78.62 MB. RACHNA did not solve any problems with 1\% tasks. The y-axis maximum is 27,600 MB (i.e., 27.6 GB).}
    \label{fig:rachna_rachnadt_heterogeneous_comms}
\end{figure}

RACHNA's and RACHNA$_{dt}$'s communication had similar trends (see Figure \ref{fig:rachna_rachnadt_heterogeneous_comms}), although RACHNA$_{dt}$ required an additional $<500$ MB. Communication increased substantially from multiple robot systems ($\leq 78.62$ MB) to collectives ($\leq 27,601.14$ MB or 27.6 GB), consistent with the $O(n^3m|S|)$ per iteration communication complexity. Communication also increased with prcent tasks and services per robot. 1000 robot collectives with ten services, five services per robot, and 50\% tasks required the most communication.

These trends were significant. RACHNA's communication increased significantly across collective sizes ($H$ ($n_{25}=71$, $n=75$) $=$ 616.84, $p<0.01$), with significant post-hoc analyses ($p<0.01$).
RACHNA's communication also grew significantly with percent tasks ($p<0.01$) and services per robot ($p<0.01$). No difference was identified between five and ten service types.

RACHNA$_{dt}$ produced similar results. Significant increases were identified across collective sizes ($H$ ($n=75$) $=$ 626.14, $p<0.01$) and percent tasks ($H$ ($n=150$) $=$ 356.24, $p<0.01$), with significant post-hoc analyses ($p<0.01$). Five and ten services per robot also differed significantly ($p<0.01$). No difference was identified between five and ten service types.

RACHNA produced near-optimal solutions for most problems (Table \ref{tab:e1_utilities_heterogeneous}), but high variability caused low worst-case utilities. Utilities were most variable with small collectives and 10\% tasks, as with homogeneous collectives.

RACHNA's utilities differed significantly across collective sizes ($H$ ($n_{25}=71$, $n=75$) $=$ 66.49, $p<0.01$). Specifically, post-hoc tests found significant differences between 25 and 50-1000 robots ($p<0.01$), 50 and 500-1000  robots ($p<0.01$), and 100 and 1000 robots ($p=0.03$). No significant differences between other collective sizes were identified. 10\% and 50\% tasks differed significantly ($p<0.01$), while no significant differences were found between five and ten services, or one and five services per robot. RACHNA$_{dt}$ outperformed RACHNA, producing optimal solutions for all trials.

\begin{table}[h!]
    \centering
      \caption{RACHNA's percent utility statistics with heterogeneous collectives.}
    \label{tab:e1_utilities_heterogeneous}
    \begin{NiceTabular}{|c|c|c||c|}
    \hline
         Service Types & Percent & Coll. & RACHNA \\\cline{4-4}
         (Per Robot) & Tasks & Size & Median (Min, Max)\\
         \hline
\multirow{10}{*}{5 (1)}&  \multirow{5}{*}{10} & 25 & 100.0 (61.29, 100.0)\\
& & 50 & 93.9 (77.42, 100.0)\\
&  & 100 &  97.56 (85.13, 100.0)\\
&  & 500 & 96.5 (94.59, 98.76)\\
&  & 1000 & 95.83 (93.13, 98.31)\\ \cline{2-4}

& \multirow{5}{*}{50} & 25  & 100.0 (99.27, 100.0)\\
& & 50 &  100.0 (99.51, 100.0)\\
&  & 100  & 99.92 (99.54, 100.0)\\
&  & 500  & 99.91 (99.8, 99.98)\\
&  & 1000 &  99.87 (99.81, 99.94)\\

\hline
\hline
\multirow{10}{*}{10 (1)} & \multirow{5}{*}{10} & 25 &  100.0 (80.43, 100.0)\\
&  & 50  & 94.57 (82.54, 100.0)\\
&  & 100 & 96.85 (87.77, 100.0)\\
&  & 500  & 96.6 (93.41, 98.64)\\
&  & 1000 & 96.26 (93.81, 97.81)\\ \cline{2-4}
& \multirow{5}{*}{50} & 25  & 100.0 (98.39, 100.0)\\
&  & 50 &  100.0 (99.5, 100.0)\\
&  & 100 &  99.92 (99.37, 100.0)\\
&  & 500 &  99.89 (99.77, 99.97)\\
&  & 1000 &  99.87 (99.8, 99.94)\\
\hline
\hline
\multirow{10}{*}{10 (5)} & \multirow{5}{*}{10} & 25  & 100.0 (73.17, 100.0)\\
&  & 50 & 97.32 (81.74, 100.0)\\
&  & 100  & 96.67 (88.99, 100.0)\\
&  & 500 & 95.52 (92.47, 98.69)\\
&  & 1000  & 96.04 (93.37, 97.69)\\\cline{2-4}
&  \multirow{5}{*}{50}& 25  & 100.0 (98.48, 100.0)\\
&  & 50  & 100.0 (99.12, 100.0)\\
&  & 100  & 99.92 (99.28, 100.0)\\
&  & 500  & 99.89 (99.71, 99.98)\\
&  & 1000  & 99.89 (99.76, 99.96)\\
\hline
    \end{NiceTabular}
  
\end{table}

\subsubsection{Simultaneous Descending Auction}

$SDA_M$'s and $SDA_{SCO}$'s results are presented separately, as they differ substantially. $SDA_M$ had a 100\% success rate.

\begin{figure}[t]
    \centering
    \includegraphics[width=\linewidth]{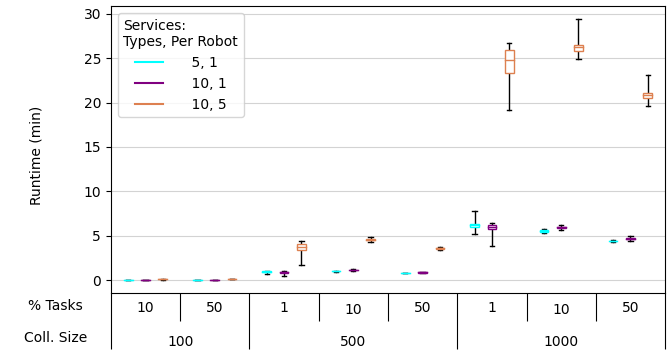}
    \caption{\textbf{$SDA_M$'s runtime (min) with heterogeneous collectives.} All multiple robot trials (i.e., 25-50 robots) completed in $<1$ s. The y-axis maximum is 30 min.}
    \label{fig:samatch_heterogeneous_runtime}
\end{figure}

 $SDA_M$'a runtimes (Figure \ref{fig:samatch_heterogeneous_runtime}) were low for multiple robot systems ($<1$ s), but increased substantially for collectives ($\leq 29$ min 29 s), consistent with its $O(mn^4)$ per iteration computational complexity. Additionally, runtimes increased with services per robot, as robots were considered for more roles. 1\% and 10\% tasks performed similarly, but had higher runtimes than 50\% tasks. The decrease in runtime can be attributed to smaller coalitions (i.e., higher percent tasks) being cheaper, thus requiring fewer bidding iterations. There was no detectable difference between five and ten service types. 
 
The increases across collective sizes ($H$ ($n=75$) $=$ 694.31, $p<0.01$) and percent tasks ($H$ ($n=150$) $=$ 24.23, $p<0.01$) were significant. Post-hoc tests revealed significant differences between 
all collective sizes ($p<0.01$), as well as 1\%-10\% and 50\% tasks ($p<0.01$); however, 1\% and 10\% tasks did not differ significantly. A significant increase between one and five services per robot was also identified ($p<0.01$). No difference between five and ten service types was found. The longest runtimes occurred with 1000 robots, ten service types, and five services per robot.

\begin{figure}[t]
    \centering
    \includegraphics[width=\linewidth]{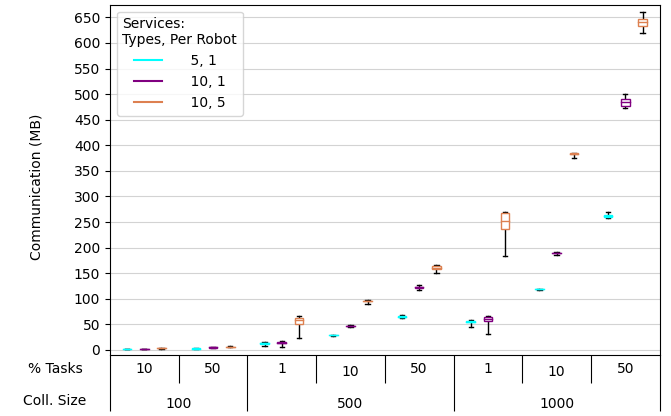}
    \caption{\textbf{$SDA_M$'s communication (MB) with heterogeneous collectives.} All multiple robot trials (i.e., 25-50 robots) required $<2$ MB. The y-axis maximum is 650 MB.}
    \label{fig:samatch_heterogeneous_comms}
\end{figure}

$SDA_M$'s total required communication (Figure \ref{fig:samatch_heterogeneous_comms}) was $<2$ MB with multiple robot systems and increased to as much as 661.39 MB with collectives. The total communication also increased with percent tasks, service types, and service types per robot, consistent with $SDA_M$'s $O((m+n)|S|)$ per iteration communication complexity. 1000 robot collectives with ten service types, five services per robot, and 50\% tasks required the most communication.

Significant differences were identified across collective sizes ($H$ ($n=75$) $=$ 693.91, $p<0.01$) and  percent tasks ($H$ ($n=150$) $=$ 153.99, $p<0.01$), with significant post-hoc analyses ($p<0.01$). The increases with service types ($p=0.04$) and services per robot ($p<0.01$) were also significant. 

$SDA_M$ derived optimal solutions 
with one service per robot, as well as 
five services per robot and 1\% or 10\% tasks. The percent utilities with five services per robot and 50\% tasks (Table \ref{tab:sa_utilities_heterogeneous}) were lower, but still near optimal. 

The difference in utility with one and five service types per robot was significant ($p<0.01$), as was the difference across collective sizes ($H$ ($n=75$) $=$ 12.39, $p=0.01$). Post-hoc tests identified significant differences between

\noindent collective sizes 25 and 500-1000 ($p<0.01$), 50 and 500-1000 ($p=0.02$ and $p=0.03$, respectively), and 100 and 500-1000 ($p=0.03$ and $p=0.04$, respectively). No significant difference was found between other collective sizes, five and ten service types, or across percent tasks.

 \begin{table}[t!]
    \centering
     \caption{\textbf{$SDA_M$'s percent utilities with heterogeneous collectives with ten service types, five services per robot, and 1\% tasks}. $SA_M$ produced optimal solutions for all other independent variable values.}
    \begin{NiceTabular}{|c||c|}
    \hline
        Coll. & Percent Utility\\
          Size & Median (Minimum, Maximum)\\
         \hline
 25  & 99.78 (98.29, 100.0)\\
50  & 99.94 (99.55, 100.0)\\
100  & 99.99 (99.87, 100.0)\\
500  & 100.0 (100.0, 100.0)\\
1000  & 100.0 (100.0, 100.0)\\
\hline
    \end{NiceTabular}
   
    \label{tab:sa_utilities_heterogeneous}
\end{table}

$SDA_{SCO}$ performed worse than $SA_M$, completing only 64.7\% (i.e., 194) of the 300 multiple robot system trials. Specifically, $SDA_{SCO}$ failed to solve any of the 75 trials with 50 robots and 10\% tasks, due to exceeding the runtime limit. Additionally, $SDA_{SCO}$ with ten service types and five services per robot exceeded the time limit for all 25 trials with 25 robots and a 10\% tasks, as well as 6 of the 25 trials with 50 robots and a 50\% tasks. The remaining multiple robot trials were completed successfully.  $SDA_{SCO}$'s evaluation did not include the 600 trials with collectives, for an overall 32.3\% success rate.

$SDA_{SCO}$'s runtimes increased with collective size, services per robot, and decreased percent tasks (Table \ref{tab:sasco_heterogeneous_runtime}). Significant differences were identified between 25 and 50 robot collectives ($p<0.01$), 10\% and 50\% tasks ($p<0.01$), and one and five services per robot ($p<0.01$) No significant difference existed between five and ten service types. These trends resulted in a maximum runtime (i.e., 1 hour 33 min) that is much too long for the near real-time allocation necessary to support high tempo applications, but short enough to support short term pre-mission planning for multiple robot systems.

\begin{table}[h]
    \centering
    \caption{\textbf{$SDA_{SCO}$'s runtimes with heterogeneous multiple robot systems.} Collective trials (i.e., 100-1000 robot)s were not attempted. No other trials were successful.}
    \label{tab:sasco_heterogeneous_runtime}
    \begin{NiceTabular}{|c|c|c|c||c|}
        \hline
       Service & Services & Percent & Coll. & Runtime (min:s:ms) \\
        Types & Per Robot & Tasks & Size & Median (Minimum, Maximum)\\
        \hline
        \hline
        \multirow{3}{*}{5} & \multirow{3}{*}{1} & 10 & 25 & 1:22:659 (0:08:932, 1:38:643)\\ \cline{3-5}
        &  & \multirow{2}{*}{50} & 25 & 0:00:172 (0:00:116, 0:00:442)\\
        & & & 50 & 0:02:979 (0:01:906, 1:54:099)\\
        \hline
        \multirow{3}{*}{10} & \multirow{3}{*}{1} & 10 & 25 & 1:26:369 (0:38:337, 1:44:126)\\
         &  & 50 & 25 & 0:02:185 (0:00:123, 0:01:124)\\ \cline{3-5}
        & & 50 & 50 & 0:02:887 (0:00:988, 0:14:159)\\
        \hline
        \multirow{2}{*}{10}& \multirow{2}{*}{5} & \multirow{2}{*}{50} & 25 &  0:05:673 (0:00:679, 16:58:993)\\
        & & & 50 & 22:50:884 (0:11:952, 93:05:483)\\
        \hline
    \end{NiceTabular}
\end{table}

The communication requirements increased significantly with collective size ($p<0.01$), as shown in Table \ref{tab:sasco_heterogeneous_comms}. As well, the amount of needed communication was significantly higher based on the percent tasks ($p<0.01$), services types ($p<0.01$), and services per robot ($p=0.04$).

\begin{table}[h]
    \centering
    \caption{\textbf{$SDA_{SCO}$'s communication with heterogeneous multiple robot systems.} Collective trials (i.e., 100-1000 robots) were not attempted. No other trials were successful.}
    \label{tab:sasco_heterogeneous_comms}
    \begin{NiceTabular}{|c|c|c|c||c|}
        \hline
       Service & Services & Percent & Coll. & Communication (MB) \\
        Types & Per Robot & Tasks & Size & Median (Minimum, Maximum)\\
        \hline
        \hline
        \multirow{3}{*}{5} & \multirow{3}{*}{1} & 10 & 25 & 50.7 (17.05, 63.1)\\\cline{3-5}
        & & \multirow{2}{*}{50} & 25 & 154.63 (101.83, 173.27)\\
        & & & 50  & 639.48 (495.05, 724.42)\\
        \hline
        \multirow{3}{*}{10} & & 10 & 25 & 77.06 (15.98, 99.72)\\\cline{3-5}
        & 1 & \multirow{2}{*}{50}& 25 & 256.22 (200.42, 325.74))\\
         & & & 50 & 1,142.34 (964.76, 1,337.88)\\
        \hline
        \multirow{2}{*}{10}& \multirow{2}{*}{5}&\multirow{2}{*}{50}  & 25 & 384.42 (306.88, 440.69.74)\\
        & & & 50 & 1,579.02 (1,535.6, 1,622.44)\\
        \hline
    \end{NiceTabular}
\end{table}

$SDA_{SCO}$ derived optimal solutions for all successful trials with five services and one service per robot, as well as all ten services and one service per robot. The median percent utilities with ten services, five services per robot, and 50\% tasks are 99.28 (minimum $=$ 96.0, maximum $=$ 100.0) with 25 robots, and 99.82 (minimum $=$ 99.27, maximum $=$ 100.0) with 50 robots. These utilities are lower than for other compositions, but still near optimal.

\subsection{Heterogeneous Collective Discussion}

The heterogeneous collective experiments assessed the algorithms' ability to produce near-optimal solutions (i.e., $>95$\% utility) for very large collectives in near real-time (i.e., $<5$ min) and using minimal communication (i.e., $<500$ MB). The \textit{hypothesis} that none of GRAPE, RACHNA, RACHNA$_{dt}$, $SDA_M$, and $SDA_{SCO}$ fully satisfied all criteria was supported.

\begin{table}[b!]
    \centering
      \caption{\textbf{Heterogeneous Results Summary: SDA$_{SCO}$ and RACHNA.} Each cell corresponds to a criterion for evaluating viability and an independent variable combination (i.e., an algorithm, services (S), services per robot (R), percent tasks (T), and a collective size). A \textcolor{ForestGreen}{\boldcheckmark} means that the algorithm met the criterion for all trials, while a \textcolor{Apricot}{\boldcheckmark} means that all trials were reasonably close to meeting the criterion. An \textcolor{red}{\textbf{X}} means that the criterion was not met, or trials were not attempted due to known algorithm limitations. A \textbf{-} means that the independent variable combination is invalid. }
    \label{tab:discussion_heterogeneous_bad}
    \resizebox{\columnwidth}{!}{
    \begin{NiceTabular}{|c|c|c?c|c|c|c|c?c|c|c|c|c|}
    \hline
        & & & \multicolumn{5}{c?}{\textbf{SDA}$_{SCO}$} & \multicolumn{5}{c|}{\textbf{RACHNA}}\\\cline{4-13}
        & & & \multicolumn{5}{c?}{Collective Size}& \multicolumn{5}{c|}{Collective Size}\\\cline{4-13}
       & \multirow{-3}{*}{\makecell{S\\R}} & \multirow{-3}{*}{\makecell{T}}  & 25 & 50 & 100 & 500 & 1000 & 25 & 50 & 100 & 500 & 1000\\
        \Xhline{2pt}
             \multirow{6}{*}{\makecell{Succ.\\Rate\\100\%}} &\multirow{3}{*}{\makecell{5/10\\1}} & 1 & \textbf{-} & \textbf{-} & \textbf{-} & \textcolor{red}{\textbf{X}}& \textcolor{red}{\textbf{X}}&\textbf{-} & \textbf{-}& \textbf{-} & \textcolor{red}{\textbf{X}}& \textcolor{red}{\textbf{X}}\\\cline{3-13}
        & & 10 & \textcolor{ForestGreen}{\boldcheckmark} & \textcolor{red}{\textbf{X}}& \textcolor{red}{\textbf{X}} & \textcolor{red}{\textbf{X}}& \textcolor{red}{\textbf{X}}&\textcolor{red}{\textbf{X}} & \textcolor{ForestGreen}{\boldcheckmark}& \textcolor{ForestGreen}{\boldcheckmark} & \textcolor{ForestGreen}{\boldcheckmark}& \textcolor{ForestGreen}{\boldcheckmark}\\\cline{3-13}
        & & 50 & \textcolor{ForestGreen}{\boldcheckmark} & \textcolor{ForestGreen}{\boldcheckmark}& \textcolor{red}{\textbf{X}} & \textcolor{red}{\textbf{X}}& \textcolor{red}{\textbf{X}}&\textcolor{ForestGreen}{\boldcheckmark} & \textcolor{ForestGreen}{\boldcheckmark}& \textcolor{ForestGreen}{\boldcheckmark} & \textcolor{ForestGreen}{\boldcheckmark}& \textcolor{ForestGreen}{\boldcheckmark}\\\cline{2-13}

           & \multirow{3}{*}{\makecell{10\\5}} & 1 & \textbf{-} & \textbf{-} & \textbf{-}& \textcolor{red}{\textbf{X}}& \textcolor{red}{\textbf{X}}& \textbf{-} & \textbf{-} & \textbf{-} & \textcolor{red}{\textbf{X}}& \textcolor{red}{\textbf{X}}\\\cline{3-13}
        & & 10 & \textcolor{red}{\textbf{X}} & \textcolor{red}{\textbf{X}}& \textcolor{red}{\textbf{X}} & \textcolor{red}{\textbf{X}}& \textcolor{red}{\textbf{X}}&  \textcolor{red}{\textbf{X}} &\textcolor{ForestGreen}{\boldcheckmark} & \textcolor{ForestGreen}{\boldcheckmark} & \textcolor{ForestGreen}{\boldcheckmark} & \textcolor{ForestGreen}{\boldcheckmark}\\\cline{3-13}
       & & 50 & \textcolor{ForestGreen}{\boldcheckmark} & \textcolor{red}{\textbf{X}}& \textcolor{red}{\textbf{X}} & \textcolor{red}{\textbf{X}}& \textcolor{red}{\textbf{X}} &\textcolor{ForestGreen}{\boldcheckmark} &\textcolor{ForestGreen}{\boldcheckmark} & \textcolor{ForestGreen}{\boldcheckmark} & \textcolor{ForestGreen}{\boldcheckmark} &\textcolor{ForestGreen}{\boldcheckmark}\\
        \Xhline{2pt}
        
        \multirow{6}{*}{\makecell{Run-\\time\\$<5$\\min}} &\multirow{3}{*}{\makecell{5/10\\1}} & 1 & \textbf{-} & \textbf{-}& \textbf{-} & \textcolor{red}{\textbf{X}}& \textcolor{red}{\textbf{X}}&\textbf{-} & \textbf{-}& \textbf{-} & \textcolor{red}{\textbf{X}}& \textcolor{red}{\textbf{X}}\\\cline{3-13}
        & & 10 & \textcolor{ForestGreen}{\boldcheckmark} & \textcolor{red}{\textbf{X}}& \textcolor{red}{\textbf{X}} & \textcolor{red}{\textbf{X}}& \textcolor{red}{\textbf{X}}&\textcolor{ForestGreen}{\boldcheckmark} & \textcolor{ForestGreen}{\boldcheckmark}& \textcolor{ForestGreen}{\boldcheckmark} & \textcolor{ForestGreen}{\boldcheckmark}& \textcolor{ForestGreen}{\boldcheckmark}\\\cline{3-13}
        & & 50 & \textcolor{ForestGreen}{\boldcheckmark} & \textcolor{ForestGreen}{\boldcheckmark}& \textcolor{red}{\textbf{X}} & \textcolor{red}{\textbf{X}}& \textcolor{red}{\textbf{X}}&\textcolor{ForestGreen}{\boldcheckmark} & \textcolor{ForestGreen}{\boldcheckmark}& \textcolor{ForestGreen}{\boldcheckmark} & \textcolor{ForestGreen}{\boldcheckmark}& \textcolor{ForestGreen}{\boldcheckmark}\\\cline{2-13}
      
           & \multirow{3}{*}{\makecell{10\\5}} & 1 & \textbf{-} & \textbf{-} & \textbf{-}& \textcolor{red}{\textbf{X}}& \textcolor{red}{\textbf{X}}& \textbf{-} & \textbf{-} & \textbf{-}& \textcolor{red}{\textbf{X}}& \textcolor{red}{\textbf{X}}\\\cline{3-13}
        & & 10 & \textcolor{red}{\textbf{X}} & \textcolor{red}{\textbf{X}}& \textcolor{red}{\textbf{X}} & \textcolor{red}{\textbf{X}}& \textcolor{red}{\textbf{X}}&  \textcolor{ForestGreen}{\boldcheckmark} &\textcolor{ForestGreen}{\boldcheckmark} & \textcolor{ForestGreen}{\boldcheckmark} & \textcolor{ForestGreen}{\boldcheckmark} & \textcolor{red}{\textbf{X}}\\\cline{3-13}
       & & 50 & \textcolor{red}{\textbf{X}} & \textcolor{red}{\textbf{X}}& \textcolor{red}{\textbf{X}} & \textcolor{red}{\textbf{X}}& \textcolor{red}{\textbf{X}} &\textcolor{ForestGreen}{\boldcheckmark} &\textcolor{ForestGreen}{\boldcheckmark} & \textcolor{ForestGreen}{\boldcheckmark} & \textcolor{ForestGreen}{\boldcheckmark} &\textcolor{red}{\textbf{X}}\\
        \Xhline{2pt}
        
            \multirow{6}{*}{\makecell{Comm.\\$<500$\\MB}} &\multirow{3}{*}{\makecell{5/10\\1}} & 1 & \textbf{-} & \textbf{-} & \textbf{-} & \textcolor{red}{\textbf{X}}& \textcolor{red}{\textbf{X}}&\textbf{-} & \textbf{-}& \textbf{-} & \textcolor{red}{\textbf{X}}& \textcolor{red}{\textbf{X}}\\\cline{3-13}
        & & 10 & \textcolor{ForestGreen}{\boldcheckmark} & \textcolor{red}{\textbf{X}}& \textcolor{red}{\textbf{X}} & \textcolor{red}{\textbf{X}}& \textcolor{red}{\textbf{X}}&\textcolor{ForestGreen}{\boldcheckmark} & \textcolor{ForestGreen}{\boldcheckmark}& \textcolor{ForestGreen}{\boldcheckmark} & \textcolor{ForestGreen}{\boldcheckmark}& \textcolor{red}{\textbf{X}}\\\cline{3-13}
        & & 50 & \textcolor{ForestGreen}{\boldcheckmark} & \textcolor{ForestGreen}{\boldcheckmark}/\textcolor{red}{\textbf{X}}& \textcolor{red}{\textbf{X}} & \textcolor{red}{\textbf{X}}& \textcolor{red}{\textbf{X}}&\textcolor{ForestGreen}{\boldcheckmark} & \textcolor{ForestGreen}{\boldcheckmark}& \textcolor{ForestGreen}{\boldcheckmark} & \textcolor{red}{\textbf{X}}& \textcolor{red}{\textbf{X}}\\\cline{2-13}

           & \multirow{3}{*}{\makecell{10\\5}} & 1 & \textbf{-} & \textbf{-} & \textbf{-}& \textcolor{red}{\textbf{X}}& \textcolor{red}{\textbf{X}}& \textbf{-} & \textbf{-} & \textbf{-} & \textcolor{red}{\textbf{X}}& \textcolor{red}{\textbf{X}}\\\cline{3-13}
        & & 10 & \textcolor{red}{\textbf{X}} & \textcolor{red}{\textbf{X}}& \textcolor{red}{\textbf{X}} & \textcolor{red}{\textbf{X}}& \textcolor{red}{\textbf{X}}&  \textcolor{ForestGreen}{\boldcheckmark} &\textcolor{ForestGreen}{\boldcheckmark} & \textcolor{ForestGreen}{\boldcheckmark} & \textcolor{red}{\textbf{X}} & \textcolor{red}{\textbf{X}}\\\cline{3-13}
       & & 50 & \textcolor{ForestGreen}{\boldcheckmark} & \textcolor{red}{\textbf{X}}& \textcolor{red}{\textbf{X}} & \textcolor{red}{\textbf{X}}& \textcolor{red}{\textbf{X}} &\textcolor{ForestGreen}{\boldcheckmark} &\textcolor{ForestGreen}{\boldcheckmark} & \textcolor{ForestGreen}{\boldcheckmark} & \textcolor{red}{\textbf{X}} &\textcolor{red}{\textbf{X}}\\
   
        \Xhline{2pt}
            \multirow{6}{*}{\makecell{Utility\\$>95$\%}} &\multirow{3}{*}{\makecell{5/10\\1}} & 1 & \textbf{-} & \textbf{-} & \textbf{-} & \textcolor{red}{\textbf{X}}& \textcolor{red}{\textbf{X}}&\textbf{-} & \textbf{-}& \textbf{-} & \textcolor{red}{\textbf{X}}& \textcolor{red}{\textbf{X}}\\\cline{3-13}
        & & 10 & \textcolor{ForestGreen}{\boldcheckmark} & \textcolor{red}{\textbf{X}}& \textcolor{red}{\textbf{X}} & \textcolor{red}{\textbf{X}}& \textcolor{red}{\textbf{X}}&\textcolor{red}{\textbf{X}} & \textcolor{red}{\textbf{X}}& \textcolor{red}{\textbf{X}} & \textcolor{Apricot}{\boldcheckmark}& \textcolor{Apricot}{\boldcheckmark}\\\cline{3-13}
        & & 50 & \textcolor{ForestGreen}{\boldcheckmark} & \textcolor{ForestGreen}{\boldcheckmark}& \textcolor{red}{\textbf{X}} & \textcolor{red}{\textbf{X}}& \textcolor{red}{\textbf{X}}&\textcolor{ForestGreen}{\boldcheckmark} & \textcolor{ForestGreen}{\boldcheckmark}& \textcolor{ForestGreen}{\boldcheckmark} & \textcolor{ForestGreen}{\boldcheckmark}& \textcolor{ForestGreen}{\boldcheckmark}\\\cline{2-13}

           & \multirow{3}{*}{\makecell{10\\5}} & 1 & \textbf{-} & \textbf{-} & \textbf{-}& \textcolor{red}{\textbf{X}}& \textcolor{red}{\textbf{X}}& \textbf{-} & \textbf{-} & \textbf{-} & \textcolor{red}{\textbf{X}}& \textcolor{red}{\textbf{X}}\\\cline{3-13}
        & & 10 & \textcolor{red}{\textbf{X}} & \textcolor{red}{\textbf{X}}& \textcolor{red}{\textbf{X}} & \textcolor{red}{\textbf{X}}& \textcolor{red}{\textbf{\textbf{X}}}&  \textcolor{red}{\textbf{X}} &\textcolor{red}{\textbf{X}} & \textcolor{red}{\textbf{X}} & \textcolor{Apricot}{\boldcheckmark} & \textcolor{Apricot}{\boldcheckmark}\\\cline{3-13}
       & & 50 & \textcolor{ForestGreen}{\boldcheckmark} & \textcolor{ForestGreen}{\boldcheckmark}& \textcolor{red}{\textbf{X}} & \textcolor{red}{\textbf{X}}& \textcolor{red}{\textbf{X}} &\textcolor{ForestGreen}{\boldcheckmark} &\textcolor{ForestGreen}{\boldcheckmark} & \textcolor{ForestGreen}{\boldcheckmark} & \textcolor{ForestGreen}{\boldcheckmark} &\textcolor{ForestGreen}{\boldcheckmark}\\
       \hline
    \end{NiceTabular}
    }
\end{table}

GRAPE was unable to address coalition formation for heterogeneous collectives, as it lacks a services model. $SDA_{SCO}$ had the next worst performance, meeting all criteria for only five of the thirty-six independent variable combinations (see Table \ref{tab:discussion_heterogeneous_bad}). Specifically, all criteria were satisfied for three independent variable combinations with five service types and one service per robot (i.e., 10\% tasks and 25 robots, 50\% tasks and 25-50 robots), two with ten service types and one service per robot (i.e., 10\%-50\% tasks and 25 robots), and none with ten service types and one five services per robot. Relaxing the criteria to $<10$ min, $<600$ MB, and $<90$\% utility does not enable $SDA_{SCO}$ to satisfy the criteria for any additional variable combinations.

$SDA_{SCO}$'s poor performance with multiple robot systems (i.e., 25-50 robots) demonstrated that the algorithm  will not scale to collectives. $SD$-$A_{SCO}$'s worst case runtimes are far too long for near real-time domains and are expected to increase significantly with collectives. Additionally, $SDA_{SCO}$'s $O(nm^ku_{max}/\epsilon_{dec})$ computational complexity made $SDA_{SCO}$ very sensitive to variations in coalition size $k$, resulting large variations in runtimes. High and variable runtimes, coupled with high communication complexities, prevent $SDA_{SCO}$ from scaling to collectives.

RACHNA satisfied all criteria for nine independent variable combinations (i.e., 50\% tasks and 25-100 robots, regardless of the collective composition). Relaxing the percent utility criteria to $>90$\% enables RACHNA to meet all criteria for two additional independent variable combination (i.e., 10\% tasks, 500 robots, five or ten service types, and one service per robot). Relaxing the runtime and communication requirements to $<10$ min and $<600$ MB respectively, does not increase the number of independent variable combinations for which RACHNA satisfies all criteria. RACHNA's primary limitations were the same as with homogeneous collectives, namely high runtimes and communication with 500-1000 robots, as well as an inability to assign coalitions to tasks requiring more robots than their utilities support.

RACHNA$_{dt}$ performed much better than RACHNA, satisfying all criteria for twenty-six independent variable combinations, as shown in Table \ref{tab:discussion_heterogeneous_good}. These independent variable combinations included all those with multiple robot systems, as well as all those with 100 robots. Additionally, RACHNA$_{dt}$ with five or ten service types and one service per robot satisfied all criteria with 1\% tasks and 50-1000 robots, as well as 10\% tasks and 500 robots. RACHNA$_{dt}$ also satisfied all criteria with ten service types, five services per robot, 1\% tasks, and 500 robots. Relaxing the criteria does not enable RACHNA$_{dt}$ to satisfy all criteria for any additional variable combinations.

RACHNA$_{dt}$'s primary strengths are its high success rates and percent utilities, which, combined with RACHNA$_{dt}$'s low runtimes with one service per robot, mean that the algorithm can be suitable for near real-time domains, given sufficient communication; however, deployed networks will necessitate reducing the communication requirement. Additionally, RACHNA$_{dt}$'s runtimes and communication with five services per robot are not suitable for near real-time domains, especially with 1000 robots.

\begin{table}[h!]
    \centering
      \caption{\textbf{Heterogeneous Results Summary: RACHNA$_{dt}$ and SDA$_M$.} Each cell corresponds to a criterion for evaluating viability and an independent variable combination (i.e., an algorithm, services (S), services per robot (R), percent tasks (T), and a collective size). A \textcolor{ForestGreen}{\boldcheckmark} means that the algorithm met the criterion for all trials, while a \textcolor{Apricot}{\boldcheckmark} means that all trials were reasonably close to meeting the criterion. An \textcolor{red}{\textbf{X}} means that the criterion was not met, or trials were not attempted due to known algorithm limitations. A \textbf{-} means that the independent variable combination is invalid.}
    \label{tab:discussion_heterogeneous_good}
    \resizebox{\columnwidth}{!}{
    \begin{NiceTabular}{|c|c|c?c|c|c|c|c?c|c|c|c|c|}
    \hline
        & & & \multicolumn{5}{c?}{\textbf{RACHNA$_{dt}$}} & \multicolumn{5}{c|}{\textbf{SDA$_M$}}\\\cline{4-13}
        & & & \multicolumn{5}{c?}{Collective Size}& \multicolumn{5}{c|}{Collective Size}\\\cline{4-13}
       & \multirow{-3}{*}{\makecell{S\\R}} & \multirow{-3}{*}{\makecell{T}}  & 25 & 50 & 100 & 500 & 1000 & 25 & 50 & 100 & 500 & 1000\\
       \Xhline{2pt}
         \multirow{6}{*}{\makecell{Succ.\\Rate\\100\%}} &\multirow{3}{*}{\makecell{5/10\\1}} & 1 & \textbf{-} & \textbf{-}& \textbf{-} & \textcolor{ForestGreen}{\boldcheckmark}& \textcolor{ForestGreen}{\boldcheckmark} & \textbf{-} & \textbf{-}& \textbf{-} & \textcolor{ForestGreen}{\boldcheckmark} & \textcolor{ForestGreen}{\boldcheckmark}\\\cline{3-13}
        & & 10 & \textcolor{ForestGreen}{\boldcheckmark} & \textcolor{ForestGreen}{\boldcheckmark}& \textcolor{ForestGreen}{\boldcheckmark} & \textcolor{ForestGreen}{\boldcheckmark}& \textcolor{ForestGreen}{\boldcheckmark} & \textcolor{ForestGreen}{\boldcheckmark}& \textcolor{ForestGreen}{\boldcheckmark} & \textcolor{ForestGreen}{\boldcheckmark} &  \textcolor{ForestGreen}{\boldcheckmark}&  \textcolor{ForestGreen}{\boldcheckmark}\\\cline{3-13}
        & & 50 & \textcolor{ForestGreen}{\boldcheckmark} & \textcolor{ForestGreen}{\boldcheckmark}& \textcolor{ForestGreen}{\boldcheckmark} & \textcolor{ForestGreen}{\boldcheckmark}& \textcolor{ForestGreen}{\boldcheckmark} & \textcolor{ForestGreen}{\boldcheckmark} & \textcolor{ForestGreen}{\boldcheckmark}& \textcolor{ForestGreen}{\boldcheckmark} & \textcolor{ForestGreen}{\boldcheckmark}& \textcolor{ForestGreen}{\boldcheckmark}\\\cline{2-13}

           & \multirow{3}{*}{\makecell{10\\5}} & 1 & \textbf{-} & \textbf{-} & \textbf{-}& \textcolor{ForestGreen}{\boldcheckmark}& \textcolor{ForestGreen}{\boldcheckmark}& \textbf{-} & \textbf{-} & \textbf{-} & \textcolor{ForestGreen}{\boldcheckmark}& \textcolor{ForestGreen}{\boldcheckmark}\\\cline{3-13}
        & & 10 & \textcolor{ForestGreen}{\boldcheckmark} & \textcolor{ForestGreen}{\boldcheckmark}& \textcolor{ForestGreen}{\boldcheckmark} & \textcolor{ForestGreen}{\boldcheckmark}& \textcolor{ForestGreen}{\boldcheckmark} & \textcolor{ForestGreen}{\boldcheckmark} & \textcolor{ForestGreen}{\boldcheckmark}& \textcolor{ForestGreen}{\boldcheckmark} & \textcolor{ForestGreen}{\boldcheckmark}& \textcolor{ForestGreen}{\boldcheckmark}\\\cline{3-13}
       & & 50 & \textcolor{ForestGreen}{\boldcheckmark} & \textcolor{ForestGreen}{\boldcheckmark}& \textcolor{ForestGreen}{\boldcheckmark} & \textcolor{ForestGreen}{\boldcheckmark}& \textcolor{ForestGreen}{\boldcheckmark} & \textcolor{ForestGreen}{\boldcheckmark} & \textcolor{ForestGreen}{\boldcheckmark}& \textcolor{ForestGreen}{\boldcheckmark} & \textcolor{ForestGreen}{\boldcheckmark}& \textcolor{ForestGreen}{\boldcheckmark}\\

       \Xhline{2pt}
        
        \multirow{6}{*}{\makecell{Run-\\time\\$<5$\\min}} &\multirow{3}{*}{\makecell{5/10\\1}} & 1 & \textbf{-} & \textbf{-} & \textbf{-} & \textcolor{ForestGreen}{\boldcheckmark} & \textcolor{ForestGreen}{\boldcheckmark} & \textbf{-} & \textbf{-} & \textbf{-} & \textcolor{ForestGreen}{\boldcheckmark} & \textcolor{Apricot}{\boldcheckmark}\\\cline{3-13}
        & & 10 & \textcolor{ForestGreen}{\boldcheckmark} & \textcolor{ForestGreen}{\boldcheckmark} & \textcolor{ForestGreen}{\boldcheckmark} & \textcolor{ForestGreen}{\boldcheckmark} & \textcolor{ForestGreen}{\boldcheckmark} & \textcolor{ForestGreen}{\boldcheckmark} & \textcolor{ForestGreen}{\boldcheckmark} & \textcolor{ForestGreen}{\boldcheckmark} & \textcolor{ForestGreen}{\boldcheckmark} & \textcolor{Apricot}{\boldcheckmark} \\\cline{3-13}
        & & 50 & \textcolor{ForestGreen}{\boldcheckmark} & \textcolor{ForestGreen}{\boldcheckmark}& \textcolor{ForestGreen}{\boldcheckmark} & \textcolor{ForestGreen}{\boldcheckmark}& \textcolor{ForestGreen}{\boldcheckmark}   & \textcolor{ForestGreen}{\boldcheckmark} & \textcolor{ForestGreen}{\boldcheckmark} & \textcolor{ForestGreen}{\boldcheckmark} & \textcolor{ForestGreen}{\boldcheckmark} & \textcolor{ForestGreen}{\boldcheckmark} \\\cline{2-13}

           & \multirow{3}{*}{\makecell{10\\5}} & 1 & \textbf{-} & \textbf{-} & \textbf{-}& \textcolor{ForestGreen}{\boldcheckmark}& \textcolor{ForestGreen}{\boldcheckmark}& \textbf{-} & \textbf{-} & \textbf{-} & \textcolor{ForestGreen}{\boldcheckmark} & \textcolor{red}{\textbf{X}}\\\cline{3-13}
        & & 10 & \textcolor{ForestGreen}{\boldcheckmark} & \textcolor{ForestGreen}{\boldcheckmark}& \textcolor{ForestGreen}{\boldcheckmark} & \textcolor{ForestGreen}{\boldcheckmark}& \textcolor{red}{\textbf{X}} &\textcolor{ForestGreen}{\boldcheckmark} & \textcolor{ForestGreen}{\boldcheckmark}& \textcolor{ForestGreen}{\boldcheckmark} &\textcolor{ForestGreen}{\boldcheckmark} & \textcolor{red}{\textbf{X}}\\\cline{3-13}
       & & 50 & \textcolor{ForestGreen}{\boldcheckmark} & \textcolor{ForestGreen}{\boldcheckmark}& \textcolor{ForestGreen}{\boldcheckmark} & \textcolor{ForestGreen}{\boldcheckmark}& \textcolor{red}{\textbf{X}} &\textcolor{ForestGreen}{\boldcheckmark} & \textcolor{ForestGreen}{\boldcheckmark}& \textcolor{ForestGreen}{\boldcheckmark} &\textcolor{ForestGreen}{\boldcheckmark} & \textcolor{red}{\textbf{X}}\\
       \Xhline{2pt}
        
                    \multirow{6}{*}{\makecell{Comm.\\$<500$\\MB}} &\multirow{3}{*}{\makecell{5/10\\1}} & 1 & \textbf{-} & \textbf{-} & \textbf{-} & \textcolor{ForestGreen}{\boldcheckmark}& \textcolor{ForestGreen}{\boldcheckmark}&\textbf{-} & \textbf{-}& \textbf{-} & \textcolor{ForestGreen}{\boldcheckmark} & \textcolor{ForestGreen}{\boldcheckmark} \\\cline{3-13}
        & & 10 & \textcolor{ForestGreen}{\boldcheckmark} & \textcolor{ForestGreen}{\boldcheckmark} & \textcolor{ForestGreen}{\boldcheckmark} & \textcolor{ForestGreen}{\boldcheckmark} & \textcolor{red}{\textbf{X}}& \textcolor{ForestGreen}{\boldcheckmark}& \textcolor{ForestGreen}{\boldcheckmark}& \textcolor{ForestGreen}{\boldcheckmark}& \textcolor{ForestGreen}{\boldcheckmark}& \textcolor{ForestGreen}{\boldcheckmark}\\\cline{3-13}
        & & 50 & \textcolor{ForestGreen}{\boldcheckmark} & \textcolor{ForestGreen}{\boldcheckmark} & \textcolor{ForestGreen}{\boldcheckmark} & \textcolor{red}{\textbf{X}} & \textcolor{red}{\textbf{X}}& \textcolor{ForestGreen}{\boldcheckmark}& \textcolor{ForestGreen}{\boldcheckmark}& \textcolor{ForestGreen}{\boldcheckmark}& \textcolor{ForestGreen}{\boldcheckmark}& \textcolor{ForestGreen}{\boldcheckmark}/\textcolor{Apricot}{\boldcheckmark}\\\cline{2-13}
      
           & \multirow{3}{*}{\makecell{10\\5}} & 1 &\textbf{-} & \textbf{-} & \textbf{-} & \textcolor{ForestGreen}{\boldcheckmark}& \textcolor{ForestGreen}{\boldcheckmark}&\textbf{-} & \textbf{-}& \textbf{-} & \textcolor{ForestGreen}{\boldcheckmark}& \textcolor{ForestGreen}{\boldcheckmark}\\\cline{3-13}
        & & 10 & \textcolor{ForestGreen}{\boldcheckmark} & \textcolor{ForestGreen}{\boldcheckmark} & \textcolor{ForestGreen}{\boldcheckmark} & \textcolor{red}{\textbf{X}} & \textcolor{red}{\textbf{X}}& \textcolor{ForestGreen}{\boldcheckmark}& \textcolor{ForestGreen}{\boldcheckmark}& \textcolor{ForestGreen}{\boldcheckmark}& \textcolor{ForestGreen}{\boldcheckmark}& \textcolor{ForestGreen}{\boldcheckmark}\\\cline{3-13}
       & & 50 & \textcolor{ForestGreen}{\boldcheckmark} & \textcolor{ForestGreen}{\boldcheckmark} & \textcolor{ForestGreen}{\boldcheckmark} & \textcolor{red}{\textbf{X}} & \textcolor{red}{\textbf{X}}& \textcolor{ForestGreen}{\boldcheckmark}& \textcolor{ForestGreen}{\boldcheckmark}& \textcolor{ForestGreen}{\boldcheckmark}& \textcolor{ForestGreen}{\boldcheckmark} & \textcolor{red}{\textbf{X}}\\
       \Xhline{2pt}
            \multirow{6}{*}{\makecell{Utility\\$>95$\%}} &\multirow{3}{*}{\makecell{5/10\\1}} & 1 & \textbf{-} & \textbf{-} & \textbf{-} &  \textcolor{ForestGreen}{\boldcheckmark}&  \textcolor{ForestGreen}{\boldcheckmark}&\textbf{-} & \textbf{-}& \textbf{-} & \textcolor{ForestGreen}{\boldcheckmark}&  \textcolor{ForestGreen}{\boldcheckmark}\\\cline{3-13}
        & & 10 &  \textcolor{ForestGreen}{\boldcheckmark}& \textcolor{ForestGreen}{\boldcheckmark}& \textcolor{ForestGreen}{\boldcheckmark}& \textcolor{ForestGreen}{\boldcheckmark}& \textcolor{ForestGreen}{\boldcheckmark}& \textcolor{ForestGreen}{\boldcheckmark}& \textcolor{ForestGreen}{\boldcheckmark}& \textcolor{ForestGreen}{\boldcheckmark}& \textcolor{ForestGreen}{\boldcheckmark}& \textcolor{ForestGreen}{\boldcheckmark}\\\cline{3-13}
        & & 50 & \textcolor{ForestGreen}{\boldcheckmark}& \textcolor{ForestGreen}{\boldcheckmark}& \textcolor{ForestGreen}{\boldcheckmark}& \textcolor{ForestGreen}{\boldcheckmark}& \textcolor{ForestGreen}{\boldcheckmark}& \textcolor{ForestGreen}{\boldcheckmark}& \textcolor{ForestGreen}{\boldcheckmark}& \textcolor{ForestGreen}{\boldcheckmark}& \textcolor{ForestGreen}{\boldcheckmark}& \textcolor{ForestGreen}{\boldcheckmark}\\\cline{2-13}
      
           & \multirow{3}{*}{\makecell{10\\5}} & 1 & \textbf{-} & \textbf{-} & \textbf{-}&  \textcolor{ForestGreen}{\boldcheckmark}&  \textcolor{ForestGreen}{\boldcheckmark}& \textbf{-} & \textbf{-} & \textbf{-} & \textcolor{ForestGreen}{\boldcheckmark}&  \textcolor{ForestGreen}{\boldcheckmark}\\\cline{3-13}
        & & 10 & \textcolor{ForestGreen}{\boldcheckmark}& \textcolor{ForestGreen}{\boldcheckmark}& \textcolor{ForestGreen}{\boldcheckmark}& \textcolor{ForestGreen}{\boldcheckmark}& \textcolor{ForestGreen}{\boldcheckmark}& \textcolor{ForestGreen}{\boldcheckmark}& \textcolor{ForestGreen}{\boldcheckmark}& \textcolor{ForestGreen}{\boldcheckmark}& \textcolor{ForestGreen}{\boldcheckmark}& \textcolor{ForestGreen}{\boldcheckmark}\\\cline{3-13}
       & & 50 & \textcolor{ForestGreen}{\boldcheckmark}& \textcolor{ForestGreen}{\boldcheckmark}& \textcolor{ForestGreen}{\boldcheckmark}& \textcolor{ForestGreen}{\boldcheckmark}& \textcolor{ForestGreen}{\boldcheckmark}& \textcolor{ForestGreen}{\boldcheckmark}& \textcolor{ForestGreen}{\boldcheckmark}& \textcolor{ForestGreen}{\boldcheckmark}& \textcolor{ForestGreen}{\boldcheckmark}& \textcolor{ForestGreen}{\boldcheckmark}\\
       \hline
    \end{NiceTabular}
    }
\end{table}

$SDA_M$ was the best performing algorithm, meeting all criteria for twenty-nine independent variable combinations (Table \ref{tab:discussion_heterogeneous_good}). Relaxing the runtime criterion to $<10$ min and the communication to $<600$ MB enabled $SDA_M$ to satisfy all criteria for four additional trials. The three independent variable combinations for which $SDA_M$ did not satisfy even the relaxed criteria were ten service types, five services per robot 1\%-50\% tasks, and 1000 robots.

Overall, $SDA_M$ met the near real-time domains' coalition formation needs reasonably well for collectives with one service per robot. Additionally, RACHNA$_{dt}$ was suitable for collectives with one service per robot, given sufficient communication. However, no algorithm fully satisfied the criteria for very large collectives (i.e., 1000 robots) with five services per robot.

\section{Discussion}

Coalition formation is important for enabling robotic collectives in applications, such as military and disaster response, to perform tasks efficiently. These applications are high-tempo and require operating in highly dynamic domains; thus, high-quality coalition formation solutions must be derived in near real-time (i.e., $<5$ min). 
The applications also typically rely on distributed, low-bandwidth networks, necessitating the use of minimal, distributed communication. A viable collective coalition formation algorithm will be required to satisfy these requirements for very large collectives, across a range of tasks.

This manuscript surveyed existing multiple robot coalition formation algorithms and found that no algorithm was known to scale to very large collectives. Some algorithms were poorly suited to collectives, due to centralization or limited solution spaces, while others' viability was unknown, due to evaluations that considered only $<100$ robots or did not consider communication. 

Auctions and hedonic games were identified as the most likely to be viable. Some auctions are decentralized and consider heterogeneous systems but had not previously been evaluated with large collectives. Meanwhile, some hedonic games can be distributed and were shown to produce solutions for multiple robot systems in real-time, but did not incorporate a services model. 

A simulation-based evaluation assessed auctions' and hedonic games' viability. The auction-based RACHNA, RACHNA$_{dt}$, $SDA_M$, and $SDA_{SCO}$ algorithms were selected due to their common auction protocols. The hedonic game-based GRAPE algorithm was selected, as it is distributed and most compatible with the services model. The hypothesis that none of the algorithms fully satisfy the coalition formation evaluation criteria (i.e., 100\% success rate, $<5$ min runtimes, $<500$ MB total communication, $>95\%$ utilities) with very large collectives was supported, as shown in Table \ref{tab:discussion}. In fact, there were collective compositions for which no algorithm was viable.

\begin{table}[t]
    \centering
    \caption{\textbf{Results Summary Table.} A \textcolor{ForestGreen}{\boldcheckmark} means that a viability criterion was satisfied for $>90$\% of homogeneous (H$_{mo}$.) or heterogeneous (H$_{tr}$) independent variable combinations. A \textcolor{Apricot}{\boldcheckmark} means that a criterion was satisfied for $>80$\%, while an \textcolor{red}{\textbf{X}} means that a criterion was satisfied for $\leq80$\%. Only independent variable combinations with collectives (i.e., $>50$ robots) are considered. Independent variable combinations for which an algorithm satisfied a relaxed constraint are counted as 1/2.}
    \label{tab:discussion}
      \resizebox{\columnwidth}{!}{
    \begin{NiceTabular}{|c?c|c?c|c?c|c?c|c?c|c|}
    \hline
       & \multicolumn{2}{c|}{\textbf{SDA$_{SCO}$}} & \multicolumn{2}{c|}{\textbf{RACHNA}}  & \multicolumn{2}{c|}{\textbf{RACHNA$_{dt}$}} &\multicolumn{2}{c|}{\textbf{GRAPE}} & \multicolumn{2}{c|}{\textbf{SDA$_M$}} \\\cline{2-11}
       & H$_{mo}$ & H$_{tr}$ & H$_{mo}$ & H$_{tr}$ & H$_{mo}$ & H$_{tr}$ &H$_{mo}$ & H$_{tr}$ &H$_{mo}$ & H$_{tr}$\\
       \Xhline{2pt}
       \multirow{4}{*}{\makecell{Succ.\\Rate\\100\%}} & \multirow{4}{*}{\textcolor{red}{\textbf{X}}} & \multirow{4}{*}{\textcolor{red}{\textbf{X}}} & \multirow{4}{*}{\textcolor{red}{\textbf{X}}} & \multirow{4}{*}{\textcolor{red}{\textbf{X}}} &
       \multirow{4}{*}{\textcolor{ForestGreen}{\boldcheckmark}}&  \multirow{4}{*}{\textcolor{ForestGreen}{\boldcheckmark}}& \multirow{4}{*}{\textcolor{red}{\textbf{X}}} & \multirow{4}{*}{\textcolor{red}{\textbf{X}}} &   \multirow{4}{*}{\textcolor{ForestGreen}{\boldcheckmark}}&  \multirow{4}{*}{\textcolor{ForestGreen}{\boldcheckmark}} \\
       &&&&&&&&&&\\
       &&&&&&&&&&\\
       &&&&&&&&&&\\
       \Xhline{2pt}
       \multirow{4}{*}{\makecell{Run-\\time\\$<5$\\min}}& \multirow{4}{*}{\textcolor{red}{\textbf{X}}} & \multirow{4}{*}{\textcolor{red}{\textbf{X}}}& \multirow{4}{*}{\textcolor{red}{\textbf{X}}} & \multirow{4}{*}{\textcolor{red}{\textbf{X}}}& \multirow{4}{*}{\textcolor{red}{\textbf{X}}} & \multirow{4}{*}{\textcolor{ForestGreen}{\boldcheckmark}}&  \multirow{4}{*}{\textcolor{red}{\textbf{X}}} & \multirow{4}{*}{\textcolor{red}{\textbf{X}}}& \multirow{4}{*}{\textcolor{red}{\textbf{X}}} & \multirow{4}{*}{\textcolor{Apricot}{\boldcheckmark}}\\
       &&&&&&&&&&\\
       &&&&&&&&&&\\
       &&&&&&&&&&\\
       \Xhline{2pt}
       \multirow{4}{*}{\makecell{Comm.\\$<500$\\MB}}  & \multirow{4}{*}{\textcolor{red}{\textbf{X}}} & \multirow{4}{*}{\textcolor{red}{\textbf{X}}}& \multirow{4}{*}{\textcolor{red}{\textbf{X}}} & \multirow{4}{*}{\textcolor{red}{\textbf{X}}}& \multirow{4}{*}{\textcolor{red}{\textbf{X}}} & \multirow{4}{*}{\textcolor{red}{\textbf{X}}}& \multirow{4}{*}{\textcolor{red}{\textbf{X}}} & \multirow{4}{*}{\textcolor{red}{\textbf{X}}}& \multirow{4}{*}{\textcolor{ForestGreen}{\boldcheckmark}} & \multirow{4}{*}{\textcolor{ForestGreen}{\boldcheckmark}}\\
       &&&&&&&&&&\\
       &&&&&&&&&&\\
       &&&&&&&&&&\\
       \Xhline{2pt}
       \multirow{4}{*}{\makecell{Utility\\$>95$\%}} &  \multirow{4}{*}{\textcolor{red}{\textbf{X}}} & \multirow{4}{*}{\textcolor{red}{\textbf{X}}} & \multirow{4}{*}{\textcolor{red}{\textbf{X}}} & \multirow{4}{*}{\textcolor{red}{\textbf{X}}} &
       \multirow{4}{*}{\textcolor{ForestGreen}{\boldcheckmark}}&  \multirow{4}{*}{\textcolor{ForestGreen}{\boldcheckmark}}& \multirow{4}{*}{\textcolor{red}{\textbf{X}}} & \multirow{4}{*}{\textcolor{red}{\textbf{X}}} &   \multirow{4}{*}{\textcolor{ForestGreen}{\boldcheckmark}}&  \multirow{4}{*}{\textcolor{ForestGreen}{\boldcheckmark}}\\
       &&&&&&&&&&\\
       &&&&&&&&&&\\
       &&&&&&&&&&\\
       \hline
    \end{NiceTabular}}
    
\end{table}

Recall that $SDA_{SCO}$ was unable to satisfy many of the performance criteria, even with smaller multiple robot systems. $SDA_{SCO}$'s inability to scale to large multiple robot systems (i.e., 50 robots) demonstrates that the algorithm will not scale to very large collectives (i.e., 1000 robots).

RACHNA's performance was also poor, due to the algorithm's inability to assign large coalitions to lower utility tasks; however, RACHNA$_{dt}$ addressed this limitation successfully. RACHNA$_{dt}$'s strength with both homogeneous and heterogeneous collectives was its high success rates and percent utilities. RACHNA$_{dt}$ satisfied the runtime criteria for $<80$\% of independent variable combinations with homogeneous collectives, but $>90$\% with heterogeneous collectives. The difference can be attributed to the number of robots that each service agent considered. Homogeneous collectives require all service agents to consider all robots, resulting in additional computation, while heterogeneous collectives only require each service agent to consider a subset of the robots. This observation is consistent with the fact that RACHNA$_{dt}$ satisfied the runtime criteria for all collectives with one service per robot and not for some collectives with five services per robot. Overall, RACHNA$_{dt}$ performed well to heterogeneous, single-service collectives, given sufficient communication. However, RACHNA$_{dt}$'s communication requirements and its runtimes with other collective compositions were excessive for near real-time domains.

GRAPE performed poorly overall, as it was only able to address homogeneous, single-service collectives. However, GRAPE, when applicable, performed well in terms of success rate, runtime, and percent utility. GRAPE has the potential to be one of the better performing algorithms, if a services model can be integrated without substantially increasing its runtimes. GRAPE's communication requirement will also need to be reduced.

$SDA_M$ was the best performing algorithm, meeting most criteria for $>90$\% of independent variable combinations. $SDA_M$'s primary limitation was its long worst case runtimes. $SDA_M$ had worse runtimes for homogeneous collectives than heterogeneous collectives; however, the underlying cause was the number of services per robot. $SDA_M$'s runtimes were longer when there were multiple services per robot, as task agents had to consider each robot for multiple roles, thus requiring more computation. The result was that $SDA_M$ performed well with homogeneous and heterogeneous collectives of single-service robots, but had runtimes that were too long with multiple-service robots.

These results were generated using a centralized simulator, in which robots' computation was performed iteratively; however, robots in real-world collectives can perform their computation in parallel. All of the algorithms' runtimes are expected to benefit from parallelization. GRAPE is expected to benefit the most, as GRAPE's computation is divided evenly among the robots, while most of the auction algorithms' computation occurs on the task and service agents, which are a small portion of the collective's agents. Parallelization is not expected to substantially impact any algorithms' success rate, communication, or solution quality. Changes to the network topology may have a more substantial impact, as the topology is known to impact GRAPE's performance \citep{Jang2018AnonymousSystem}, and such changes have not been studied for the auction-based algorithms. Spatial aspects can also impact real-world deployments. The robots' embodiment is addressed through the use of the services model, and robots' positions can be incorporated into the utilities \citep{Jang2018AnonymousSystem}. However, robots' position in the environment can impact their access to communication (e.g., limited broadcast distance, obstacles blocking network signals), which may result in lower utilities for all algorithms if robots disconnect from the network during coalition formation.

Overall, no algorithm fully satisfied the performance criteria for very large homogeneous or heterogeneous collectives. Single-service collective coalition formation can be addressed by $SDA_M$ or, given sufficient communication, GRAPE. Additionally, coalition formation for heterogeneous collectives with single-service robots can be addressed by $SDA_M$ or, given sufficient communication, RACHNA$_{dt}$. However, no algorithm successfully addresses the stated coalition formation performance criteria for homogeneous or heterogeneous collectives with multiple service robots. 

Future work must develop faster, low-communication algorithms in order to address these collective compositions. $SDA_M$ was the best-performing algorithm with other collective compositions, but it unlikely that modifying $SDA_M$ will suffice. The underlying bipartite matching algorithm is responsible for much of the computation, limiting the potential runtime gains. RACHNA$_{dt}$ is similarly limited by the the computation required to determine whether to accept bids, while $SDA_{SCO}$'s and RACHNA's performance results indicate that they are not well-suited to collectives. These results suggest that a potential path forward is to consider auctions in which computation is distributed more evenly across the collective (i.e., not combinatorial auctions). Modification of GRAPE may also be a viable path forward, if the services model can be incorporated without substantially increasing runtime. This option requires reducing GRAPE's communication; however, it is plausible that the communication reduction can be achieved by relaxing the requirement that only one robot modify the coalition structure during an iteration, thus reducing the number of times communication occurs. A successful near-real time, minimal communication coalition formation algorithm for multiple service collectives will facilitate more diverse and sophisticated collective applications.

\section{Conclusion}

Robotic collectives in the military and disaster response domains will require scalable real-time coalition formation algorithms that produce high quality solutions and use little communication. This manuscript assessed existing algorithms' applicability to homogeneous and heterogeneous collectives with up to 1000 robots and identified distributed hedonic games and auctions as the most promising algorithm types. A simulation-based evaluation of a hedonic game, GRAPE, and auction-based algorithms, RACHNA, the novel variant RACHNA$_{dt}$, $SA_M$, and $SA_{SCO}$ found that no algorithms consistently met the requirements for collective coalition formation with achievable missions. $SDA_M$, RACHNA$_{dt}$, and GRAPE each showed promise with certain collective compositions, but no algorithm fully addressed the needs of collectives with multiple service robots. Future collective coalition formation algorithms will need lower computation and communication in order to address these needs.

\backmatter

\section*{Statements and Declarations}

\bmhead{Ethics approval and consent to participate}

Not applicable.

\bmhead{Consent for publication}

Not applicable.

\bmhead{Availability of data and material}

The datasets generated during the current study are available from the corresponding author on reasonable request.

\bmhead{Funding}

Diehl has received research support from the Defense Advanced Research Projects Agency (DARPA).

\bmhead{Competing interests}

The authors have no competing interests to declare outside of funding.

\bmhead{Authors' contributions}

Both authors contributed to the study conception and design. Coding, data collection, and analysis were performed by G.D., supervised by J.A.A.. The first draft of the manuscript was written by G.D., and J.A.A. provided feedback on previous versions of the manuscript. Both authors read and approved the final manuscript.

\bmhead{Acknowledgments}

Diehl's Ph.D. was supported by an ARCS Foundation Scholar award and the Defense Advanced Research Projects Agency (DARPA). The views, opinions, and findings expressed are those of the author and are not to be interpreted as representing the official views or policies of the Department of Defense or the U.S. Government.


 \bibliography{references.bib}

\end{document}